\documentclass[epj,nopacs]{svjour}
\usepackage{graphicx}

\begin{document}

\title{\boldmath  Experimental study of the $e^+e^-\to n\bar{n}$ 
process at the VEPP-2000 $e^+e^-$ collider with the SND detector }
\authorrunning{M.~N.~Achasov et al.}
\titlerunning{Experimental study of the $e^+e^-\to n\bar{n}$
process with the SND detector}

\author{{\large The SND Collaboration}\\ \\
M.~N.~Achasov\inst{1,2} \and
A.~Yu.~Barnyakov\inst{1,3} \and
K.~I.~Beloborodov\inst{1,2} \and
A.~V.~Berdyugin\inst{1,2} \and
D.~E.~Berkaev\inst{1,2} \and
A.~G.~Bogdanchikov\inst{1} \and
A.~A.~Botov\inst{1} \and
G.~S.~Chizhik\inst{1,2} \and
T.~V.~Dimova\inst{1,2} \and
V.~P.~Druzhinin\inst{1,2} \and
L.~V.~Kardapoltsev\inst{1,2} \and
A.~G.~Kharlamov\inst{1,2} \and
V.~A.~Kladov\inst{1,2} \and
I.~A.~Koop\inst{1,2,3} \and
A.~A.~Korol\inst{1,2} \and
D.~P.~Kovrizhin\inst{1} \and
A.~S.~Kupich\inst{1,2} \and
A.~P.~Lysenko\inst{1} \and
N.~A.~Melnikova\inst{1} \and
N.~Yu.~Muchnoi\inst{1,2} \and
A.~E.~Obrazovsky\inst{1} \and
E.~V.~Pakhtusova\inst{1} \and
E.~A.~Perevedentsev\inst{1,2} \and
K.~V.~Pugachev\inst{1,2} \and
S.~I.~Serednyakov\inst{1,2}\thanks{Corresponding author: seredn@inp.nsk.su} \and
Z.~K.~Silagadze\inst{1,2} \and
P.~Yu.~Shatunov\inst{1,2} \and
Yu.~M.~Shatunov\inst{1,2} \and
D.~A.~Shtol\inst{1} \and
D.~B.~Shwartz\inst{1,2} \and
I.~K.~Surin\inst{1} \and
Yu.~V.~Usov\inst{1} \and
I.~M.~Zemlyansky\inst{1,2} \and
V.~N.~Zhabin\inst{1} \and
V.~V.~Zhulanov\inst{1,2}
}
\institute{Budker Institute of Nuclear Physics, SB RAS, Novosibirsk,
630090, Russia \and
Novosibirsk State University, Novosibirsk, 630090, Russia \and
Novosibirsk State Technical University,Novosibirsk,630073, Russia 
}

\date{ }

\abstract{
The process $e^+e^-\to n\bar{n}$ is studied in the
experiment at the VEPP-2000 $e^+e^-$  collider with the SND detector.
The technique of the time measurements in the multichannel NaI(Tl)
electromagnetic calorimeter is used to select $n\bar{n}$ events.
The value of the measured cross section in the center-of-mass energy range
from 1.894 to 2 GeV varies from 0.5 to 0.35 nb.
The effective neutron timelike form factor is derived from
the measured cross section and compared with the proton form factor. 
The ratio of the neutron electric and magnetic form
factors is obtained from the analysis of the antineutron polar
angle distribution and found to be consistent with unity. 
}
\maketitle

\section*{Introduction\label{sec:intro}}
Measurement of the $e^+e^-$ annihilation to nucleon-anti\-nuc\-leon pairs
allows to study the nucleon internal structure described by the
timelike
electromagnetic form factors,  electric $G_E$ and magnetic $G_M$.
The $n\bar{n}$ production cross section is given by the following
equation:
\begin{eqnarray}
\frac{d\sigma}{d\Omega}&=&\frac{\alpha^{2}\beta}{4s}
\bigg[ |G_M(s)|^{2}(1+\cos^2\theta)\nonumber\\
&+&\frac{1}{\gamma^2}|G_E(s)|^{2}\sin^2\theta
\bigg]
\label{eqB1}
\end{eqnarray}
where $\alpha$ is the fine structure constant, $s=4E_b^2=E^2$,
$E_b$ is the beam  energy, $E$ is the center-of-mass (c.m.)  energy,
$\beta = \sqrt{1-4m_n^2/s}$, $m_n$ is the neutron mass, $\gamma = E_b/m_n$,
and $\theta$ is the antineutron production polar angle.
The $|G_E/G_M|$ ratio can be extracted
from the analysis of the measured $\cos\theta$ distribution.
At the threshold  $|G_{E}| = |G_{M}|$.
The total cross section has the following form:
\begin{equation}
\sigma(s) =
\frac{4\pi\alpha^{2}\beta}{3s}(1+\frac{1}{2\gamma^2})|F(s)|^2,
\label{eqB2}
\end{equation}
with
\begin{equation}
|F(s)|^2=\frac{2\gamma^2|G_M(s)|^2+|G_E(s)|^2}{2\gamma^2 +1 }.
\label{eqB3}
\end{equation}
The function $F(s)$ is the so-called effective form factor, which is equal
to unity for pointlike particle. It is this function that is measured
in most of $e^+e^-\to p\bar{p}$ and $n\bar{n}$ experiments. One can see from 
Eqs.~(\ref{eqB1}) and (\ref{eqB3})  that the relative contribution
of the $|G_E(s)|^2$ term decreases with energy as $1/E_b^2$.

 The $e^+e^-\to n\bar{n}$ cross section near threshold was measured previously
in the FENICE~\cite{FENICE}
and SND~\cite{SNND} experiments.
Recently, BESIII results~\cite{BES} on the study of the $e^+e^-\to n\bar{n}$
process above 2 GeV were published.
In this work we present a new measurement of the $e^+e^-\to n\bar{n}$ cross 
section in the SND experiment.

%============================= Fig. 1 ================================
\begin{figure*}
\centering
\includegraphics [width = 0.7\textwidth]{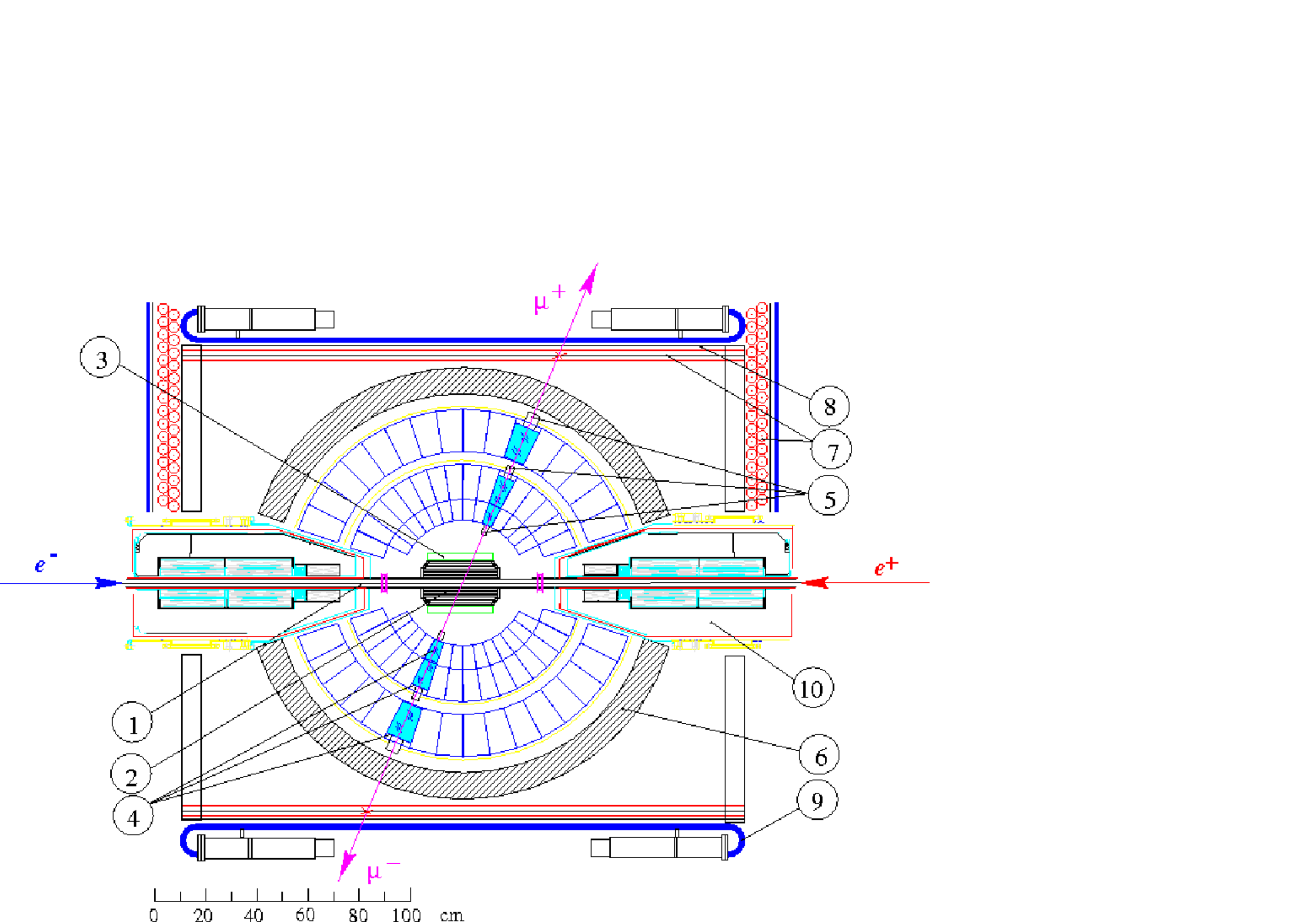}
\caption{SND detector, section along the beams: (1) beam pipe,
(2) tracking system, (3) aerogel Cherenkov counters, (4) NaI (Tl)
crystals, (5) vacuum phototriodes, (6) iron absorber, (7) proportional
tubes,
(8) iron absorber, (9) scintillation counters, (10) VEPP-2000 focusing
solenoids.}
\label{fig:sndt}
\end{figure*}
%============================= Fig. 1 ================================

\section{Collider, detector, experiment\label{sec:Exper}}
The experiment was carried out at the VEPP-2000 $e^+e^-$ 
collider~\cite{VEPP2k} with the SND 
detector~\cite{SNDet1,SNDet2,SNDet3,SNDet4}.
VEPP-2000 operates in the c.m. energy range from 0.3 to 2.0 GeV. The collider
has two collision regions, one of which is occupied by the SND detector.
The collider luminosity ranges from $10^{29}$~cm$^{-2}$s$^{-1}$ near 0.3 GeV
up to $7\times 10^{31}$~cm$^{-2}$s$^{-1}$ at the maximum energy. 
The beam energy and its spread during data taking is measured by the laser 
Compton back-scattering system~\cite{CBS}. The accuracy of the
energy measurement is 50 keV. The beam energy spread above the $n\bar{n}$ 
threshold is about 0.7 MeV.

The SND (Spherical Neutral Detector) is a general-purpose non-magnetic 
detector for a low energy collider (Fig.\ref{fig:sndt}). It
consists of a tracking system, an aerogel Cheren\-kov detector, a three-layer 
spherical NaI(Tl) electromagnetic calorimeter
(EMC) and a muon detector. The latter consists of layers of
proportional tubes and scintillation counters with an 1 cm iron sheet
between them.  The EMC is the main part of SND. It is
intended to measure the electromagnetic shower energy and angles, but is
also suitable to detect antineutrons. At the kinetic
energy of several tens of MeV the antineutron annihilation length in NaI(Tl)
does not exceed 20 cm~\cite{Annih},
which is significantly less than the EMC thickness (35 cm of
NaI(Tl))~\cite{SNND}.
This leads to a high absorption efficiency of produced
antineutrons in the SND calorimeter.

The data for this analysis were taken 
in the energy range from the $n\bar{n}$ threshold up to 2 GeV,
in 7 energy points in the 2017 run and in 7 points in the 2019 run.
The total integrated luminosity of these data is about 30 pb$^{-1}$.
The typical collider instant luminosity in the experiment
was about $2\times 10^{31}$ cm$^{-2}$s$^{-1}$.
To study background, we also analyze data with an integrated luminosity
of $20$ pb$^{-1}$ collected below the $n\bar{n}$ threshold, in the range
$E_b=900$--939 MeV. 
\section{Backgrounds and events selection \label{sec:EvSelect}}
   The background in this experiment is of three types:
physical, beam-induced, and cosmic-ray.
The physical background arises from all $e^+e^-$ annihilation
processes, in particular, those with $K_L$ meson in the final state.
The beam-induced background comes from interactions of off-energy
beam particles with elements of the collider magnetic system and the walls
of the beam pipe near the $e^+e^-$ interaction region. Beam particles can
lose energy through the radiative Bhabha scattering, beam-gas
scattering, and internal beam (Touschek) scattering. The total EMC energy
deposition in most of beam background events does not exceeds
the beam energy.
The EMC signals from physical and beam-induced background events are
synchronized with the beam revolution frequency (12.3 MHz). 
In contrast, the cosmic-ray background is evenly distributed in time. 

The $n\bar{n}$ events are very different from events of other
$e^+e^-$ annihilation processes.
Below 2 GeV the  neutron from $n\bar{n}$ pair has low energy and 
therefore gives low energy deposition in the calorimeter.  In this analysis,
the signal from neutrons is not used.
The antineutron annihilates inside the EMC and produces
pions, protons, neutrons  with the total
energy up to $2m_n$. Such an annihilation ``star'' in the EMC is a main
sign of the neutron-antineutron event. 
In SND, clusters in the calorimeter with energy deposition greater
than 20 MeV not associated with charged tracks originated from the 
interaction region are reconstructed as photons. 
Typically, a $n\bar{n}$ event looks like a
multiphoton event. A small part of the events contains off-center
tracks in the drift chamber.
In this analysis, to estimate antineutron direction we calculate
the so-called event momentum $\vec{P}_{\rm EMC}=\sum_i E_i \vec{r}_i$,
where $E_i$ is the energy deposition in EMC crystal $i$, and 
$\vec{r}_i$ is its position unit vector. The polar angle of the event
momentum ($\theta_a$) is taken as an estimate of the antineutron polar
angle.
\begin{figure*}
\includegraphics[width=0.45\textwidth]{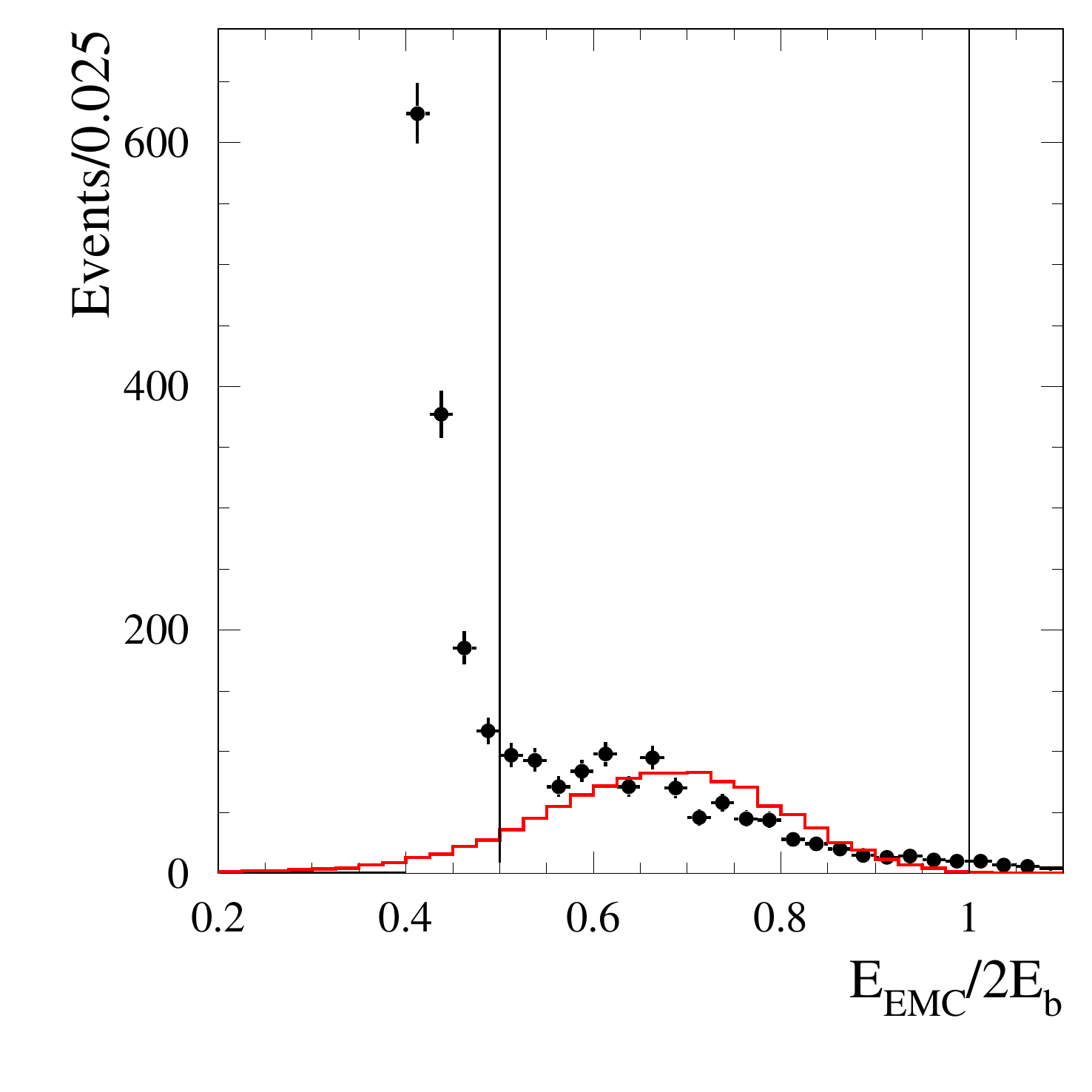} \hfill
\includegraphics[width=0.45\textwidth]{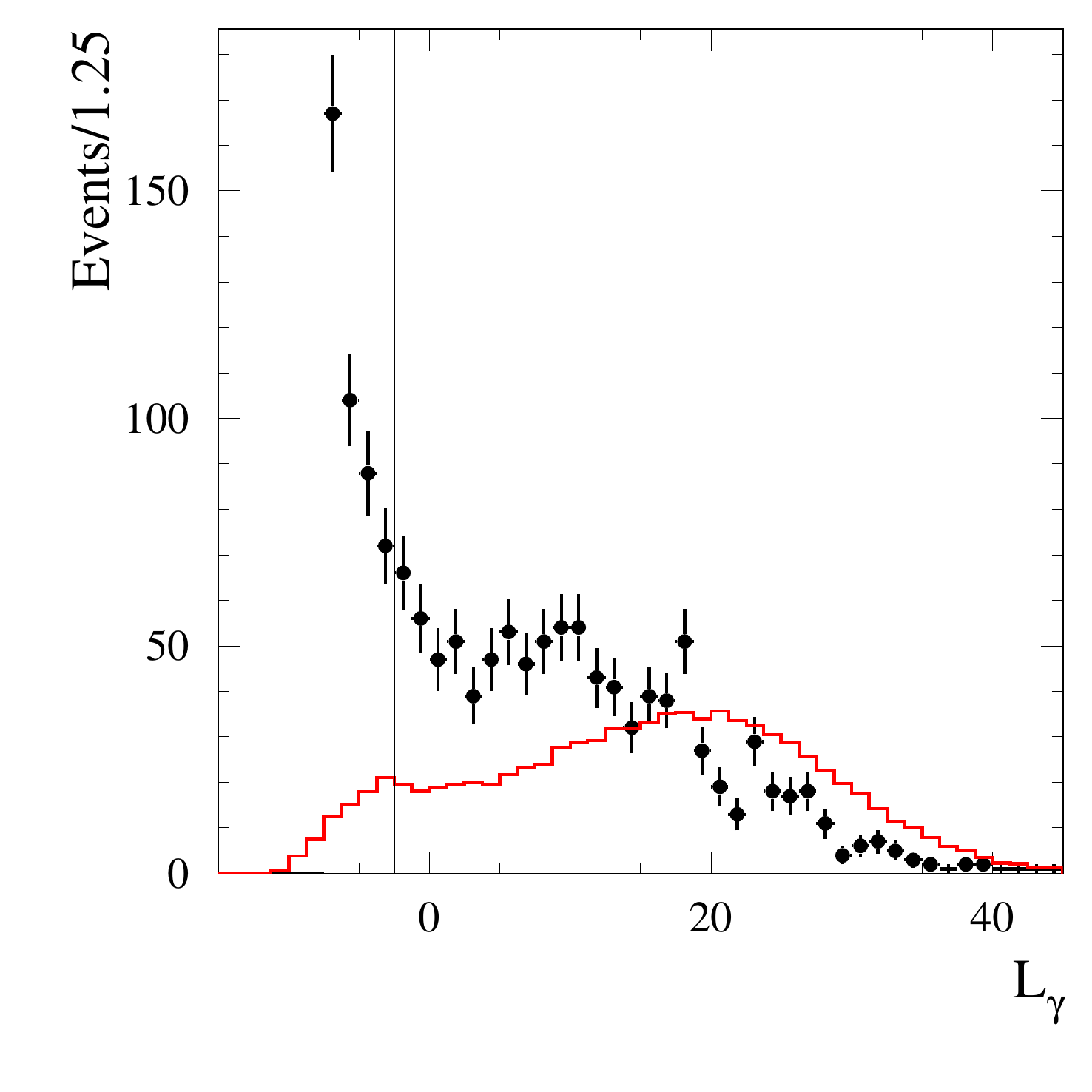} \\
\parbox[h]{0.47\textwidth}{\caption { The distribution of
the normalized total EMC energy deposition $E_{\rm EMC}/2E_b$ for 2019
data events with $E_b=945$, 950, 951 MeV (points with error bars)
Events are
satisfied the standard selection criteria except for the condition on
$E_{\rm EMC}$. The histogram represents the same distribution for simulated  
signal events. The vertical lines indicate the boundaries of the
condition $E_b<E_{\rm EMC}<2$ GeV.\label{fig:ethresh}}}\hfill
\parbox[h]{0.47\textwidth}{\caption { 
The $L_\gamma$ distribution for 2019 data events with $E_b=945$, 950,
951 MeV (points with error bars). Events are satisfied the standard selection
criteria except for the condition on $L_\gamma$.
The histogram represents the same distribution for simulated
signal events. The vertical line indicates the boundary of the
condition $L_\gamma>-2.5$.}
\label{fig:xinm}}
\end{figure*}
Basing on specific properties of signal and background events, the following 
criteria are chosen to select $n\bar{n}$ candidates.
\begin{enumerate}
\renewcommand{\labelitemi}{--}
\item No charged tracks in the drift chamber are found in an event
($n_{\rm ch}=0$).
\item The reconstructed antineutron polar angle lies in the ``large-angle''
region of the calorimeter $36^\circ<\theta_a<144^\circ$.
\item The absence of a signal in the muon system (coincidence of  
proportional tubes and scintillation counters) is required. This is the
most efficient selection condition against the cosmic-ray background.
\item 
The total energy deposition in the EMC is required to be within the limits
$E_b<E_{\rm EMC}<2E_b$ GeV. The $E_{\rm EMC}/(2E_b)$ distribution for data and 
simulated signal events is shown in Fig.\ref{fig:ethresh}. The sharp rise in the 
spectrum below $E_{\rm EMC}=E_b$ is due to the beam-induced background.
\item The large unbalanced total event momentum is measured in the
calorimeter ($P_{\rm EMC}>0.4E_{\rm EMC}$). This condition suppresses the
$e^+e^-$annihilation background.
\item The most energetic photon in an event has the polar angle in the range
$27^\circ$--$153^\circ$. This condition suppresses $e^+e^-\to \gamma\gamma$
background.
\item The transverse EMC energy profile of the most energetic photon is required
to be not consistent with the electromagnetic shower profile~\cite{xi2gam}.
The distribution of the corresponding
logarithmic likelihood function $L_\gamma$ for data and simulated signal 
events is shown in Fig.\ref{fig:xinm}. The condition $L_\gamma>-2.5$ is used.
The steep rise in the distribution at negative values is due to the $e^+e^-$
annihilation background containing real photons.
\item The cosmic-ray background is suppressed by the requirement that there
be no cosmic-ray track in the calorimeter. The cosmic-ray track is
reconstructed as a group of calorimeter crystal hits positioned along a 
straight line with $R_{\rm min} >10$ cm, where $R_{\rm min}$ is a distance 
between the track and the detector center.
\item For suppression of the cosmic-ray shower events, a special parameter has
been developed.
The moment of inertia tensor is constructed from the coordinates of the 
EMC crystals weighted by their energy depositions.
The tensor is then diagonalized, and the ratio of the 
smallest to the largest eigenvalues $R_T$ is calculated. We require 
that $R_T<0.4$ and that the distance
between the ``center of mass'' of the EMC crystals and the detector
center be greater than 10 cm.
\item The energy deposition in the third layer of the EMC $E_3<0.75E_b$.
This parameter is also used to suppress the cosmic-ray background.
\label{list:Condts}
\end{enumerate}

   As a result of applying the criteria described above, we select in
about 200 data events per pb$^{-1}$, which corresponds to a 
signal-to-background ratio of about 0.5.
\section{Determining  the number of $n\bar{n}$ events for the 2019 run 
\label{sec:Tim19}}
%============================= Fig. N ================================
\begin{figure*}
\includegraphics[width=0.45\textwidth]{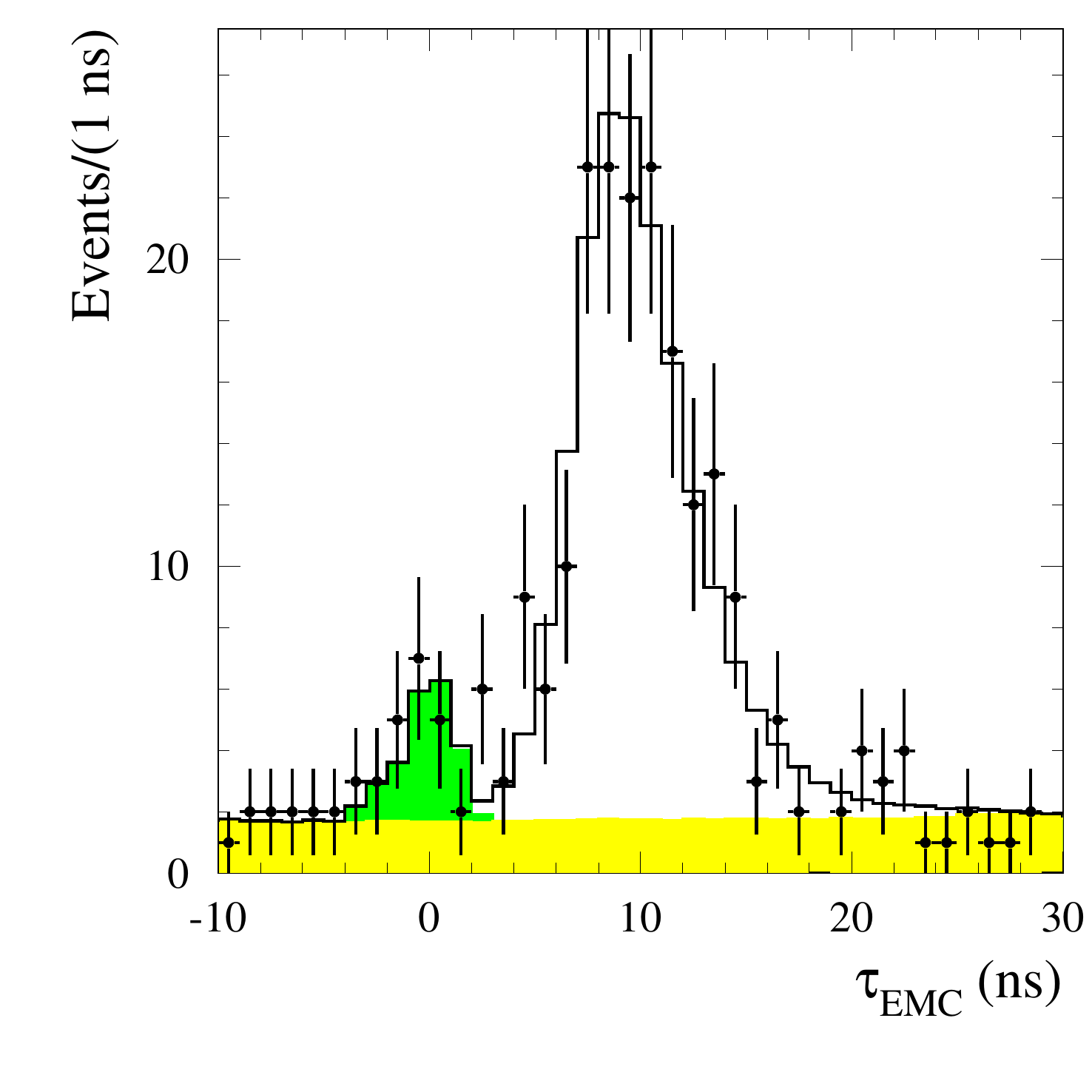} \hfill
\includegraphics[width=0.45\textwidth]{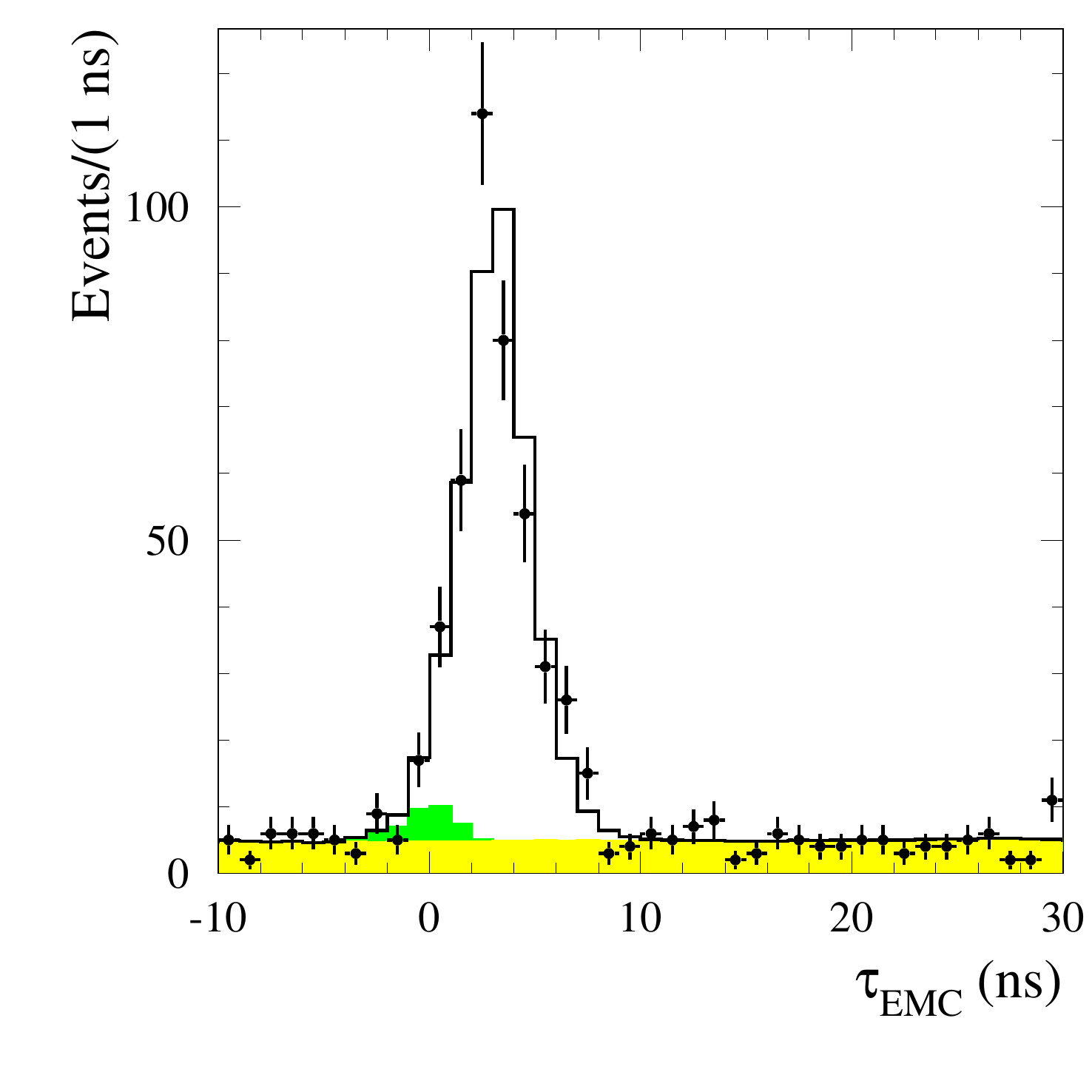} 
\caption {The $\tau_{\rm EMC}$ distribution for selected data events
collected in 2019 (points with error bars) at $E_b=945$ MeV (left panel)
and at $E_b=973$ MeV (right panel). The solid histogram is the result of
the fit described in the text. The light-shaded (yellow) histogram shows
the fitted cosmic-ray background. The medium-shaded (green) region represents
the fitted beam-induced plus physical background.\label{fig:his19}}
\end{figure*}
Due to a low antineutron velocity in the energy region under study,
its signal in the EMC is delayed with respect to the typical $e^+e^-$
annihilation event, e.g. from the process $e^+e^-\to\gamma\gamma$. This
delay is about 10 ns at $E_b=945$ MeV, and about 4 ns at 973 MeV
\footnote{Here and below in the text, we present the values of the beam energy
rounded off to an integer. More accurate energy values are given in
Table~\ref{tab:crsect}}.

In 2019, new calorimeter electronics~\cite{Timr} was installed on the SND 
detector. For each EMC crystal, the signal from the photodetector shaped with
an integration time of about 1 $\mu$s is digitized by a flash ADC
with a sampling rate of 36 MHz (three times the beam revolution
frequency). The measured signal shape is fitted
by a function previously obtained using $e^+e^- \to e^+e^-$ events. 
From the fit, the signal amplitude and arrival time are determined. 
The event time $\tau_{\rm EMC}$ is calculated as a weighted average of
EMC crystal arrival times with the energy deposition used as a weight. The
averaging is done over crystals with energy deposition of more than 25 MeV.
The time resolution measured using $e^+e^-\to \gamma\gamma$ events is about
0.7 ns.

The $\tau_{\rm EMC}$ distributions for selected data events at $E_B=945$ MeV 
and 973 MeV are shown in Fig.~\ref{fig:his19}. Time zero corresponds to the
average time for $e^+e^-\to \gamma\gamma$ events. The distribution consists
of the nearly uniform cosmic-ray distribution, the distribution for the 
beam-induced and physical backgrounds, which is peaked near zero, and the wide
delayed $n\bar{n}$ distribution. The width of the $n\bar{n}$ distribution is 
determined by the spread of the antineutron annihilation points, from the wall
of the beam pipe to the rear wall of the calorimeter.  
The distribution is fitted by  a sum of time spectra for these three
components:
\begin{equation}
F(t)=N_{n\bar{n}}H_{n\bar{n}}(t)+N_{\rm csm}H_{\rm csm}(t)+N_{\rm bkg}H_{\rm bkg}(t),
\label{eq17}
\end{equation}
where histograms  $H_{n\bar{n}}$, $H_{\rm csm}$ and $H_{\rm bkg}$  
are the $\tau_{\rm EMC}$ distributions (normalized to unity) for signal,
cosmic background, and physical + beam-induced background, respectively.
$N_{n\bar{n}}$, $N_{\rm csm}$, and $N_{\rm bkg}$ are the number of events
for these components, which are determined from the fit.

Our MC simulation reproduces the $n\bar{n}$ time distribution incorrectly. 
In particular, the time resolution is strongly underestimated in the simulation
for both $e^+e^-\to \gamma\gamma$ and $e^+e^-\to n\bar{n}$ events. From the
spread of the arrival times measured in an event in different EMC crystals, we
estimate that the time resolution for $n\bar{n}$ events is larger than that 
for $\gamma\gamma$ events by a factor of 2.4. Therefore, we convolve the 
MC time spectrum with a Gaussian function with a standard deviation of 
$\sigma_{\rm G}=1.7 \pm 0.2$ ns. The quoted uncertainty is estimated from
the simultaneous fit to the time spectra for $E_b=945$, 950, 951, and 956 MeV
with $\sigma_{\rm G}$ floating. 
%============================= Table N ================================
\begin{table*}
\centering
\caption{The beam energy ($E_b$), integrated luminosity ($L$),
number of selected $n\bar{n}$ events ($N_{n\bar{n}}$), the factor taking into
account radiative corrections and energy spread ($1+\delta$), detection
efficiency ($\varepsilon$), measured $e^+e^-\to n\bar{n}$ cross
section $\sigma$, and neutron effective form factor ($F_n$). 
The quoted errors for $N$, $\sigma$ are statistical and 
systematic. For the detection efficiency, the systematic uncertainty
is quoted. For $F_n$, the combined statistical and systematic uncertainty is
listed. Rows 1--7 (8--14) list data for 2017 (2019). 
\label{tab:crsect}}
\begin{tabular}{cccccccc}
N & $E_b$(MeV) & $L$(pb)& $N_{n\bar{n}}$ & $1+\delta$ &$\varepsilon$
&$\sigma$(nb) & $F_n$\\
\hline
 1 & 942.1  & 1.48 & $107\pm20\pm16$ & 0.697 & $0.186\pm0.021$ &
 $0.554\pm0.106\pm0.105$ &$0.454\pm0.062$ \\
 2 & 950.5  & 1.09 & $105\pm15\pm 6$ & 0.791 & $0.208\pm0.024$ &
 $0.589\pm0.084\pm0.077$ &$0.330\pm0.032$ \\
 3 & 960.8  & 1.39 & $123\pm17\pm 9$ & 0.839 & $0.203\pm0.024$ &
 $0.521\pm0.070\pm0.072$ &$0.268\pm0.026$ \\
 4 & 971.4  & 2.04 & $138\pm19\pm12$ & 0.871 & $0.202\pm0.024$ &
 $0.385\pm0.052\pm0.056$ &$0.212\pm0.021$ \\
 5 & 982.1  & 1.41 & $112\pm14\pm 9$ & 0.896 & $0.195\pm0.023$ &
 $0.453\pm0.058\pm0.065$ &$0.218\pm0.021$ \\
 6 & 991.4  & 1.39 & $ 96\pm15\pm 8$ & 0.914 & $0.191\pm0.023$ &
 $0.393\pm0.060\pm0.058$ &$0.196\pm0.021$ \\
 7 & 1003.4 & 3.22 & $188\pm22\pm19$ & 0.933 & $0.190\pm0.023$ &
 $0.329\pm0.038\pm0.052$ &$0.174\pm0.017$ \\
 8 & 944.8  & 2.59 & $159\pm14\pm 3$ & 0.745 & $0.194\pm0.017$ &
 $0.427\pm0.038\pm0.038$ &$0.334\pm0.021$ \\
 9 & 950.1  & 2.08 & $138\pm13\pm 2$ & 0.789 & $0.187\pm0.016$ &
 $0.450\pm0.042\pm0.041$ &$0.291\pm0.019$ \\
10 & 951.0  & 2.40 & $175\pm14\pm 3$ & 0.795 & $0.192\pm0.017$ &
 $0.479\pm0.039\pm0.044$ &$0.294\pm0.018$ \\
11 & 956.0  & 1.94 & $146\pm13\pm 3$ & 0.820 & $0.190\pm0.017$ & 
 $0.483\pm0.044\pm0.045$ &$0.272\pm0.018$ \\
12 & 962.7  & 2.20 & $153\pm14\pm 7$ & 0.846 & $0.186\pm0.017$ &
 $0.442\pm0.040\pm0.046$ &$0.242\pm0.017$ \\
13 & 973.0  & 4.90 & $375\pm22\pm17$ & 0.875 & $0.184\pm0.017$ &
 $0.479\pm0.028\pm0.050$ &$0.234\pm0.014$ \\
14 & 988.2  & 1.89 & $108\pm13\pm 9$ & 0.908 & $0.175\pm0.016$ &
 $0.359\pm0.042\pm0.047$ &$0.190\pm0.017$ \\
\hline
\end{tabular}
\end{table*}

It is also observed that the right 
tails of the $\tau_{\rm EMC}$ distribution in data and simulation are
different. This difference is partly explained by incorrect antineutron
annihilation cross sections used in MC simulation. 
We study antineutron annihilation in simulation using a thin absorber of 
different materials, and compare the extracted cross sections with those 
measured in Ref.~\cite{Annih}. It is found that the simulation underestimates
the annihilation cross section. The difference with experiment is greater for
materials with higher atomic number ($A$). For NaI, the antineutron annihilation
length calculated from the results of Ref.~\cite{Annih} is 7.7 (16.7) cm at 
$E_b=945 (990)$ MeV. In simulation, it is greater by a factor of 1.7 (1.2),
respectively. For a lower-$A$ material, such as aerogel, the same scale
factor is 1.3 (1.05). Using the information about the position of the
antineutron annihilation point and the scale factors defined above
we reweight simulated events. It is assumed that antineutron elastic scattering,
which effectively reduces the annihilation length, is simulated
correctly. With the time distribution obtained using reweighed simulated events
the fit is much better, but not satisfactory.   
To improve the fit quality, we modify the simulated distribution 
as follows
\begin{eqnarray}
H_{n\bar{n}}(t)&=&(1-\kappa)H_1^{\rm MC}(t)+\kappa H_2(t),\nonumber\\
H_2(t)&=&H_2^{\rm MC}(t)w(t)\Big/\int H_2^{\rm MC}(t)w(t)dt,\label{alen}
\end{eqnarray}
where $H_1^{\rm MC}(t)$ and $H_2^{\rm MC}(t)$ are
the simulated distributions for events, in which the antineutron
annihilates before and in the calorimeter, respectively,
$\kappa$ is the fraction of events with the antineutron annihilation in
the calorimeter. The distributions $H_1^{\rm MC}(t)$ and $H_2^{\rm
MC}(t)$ are normalized to unity. The weight $w(t)$ is calculated as follows
\begin{equation}
w(t)=\exp{(-\alpha_n\beta c t)},
\end{equation}
where $\beta c$ is the antineutron velocity, and the parameter 
$\alpha_n$ is floating in the fit. 

The shape of the physical + beam-induced background $H_{\rm bkg}$ is 
measured at energies below the $n\bar{n}$ threshold (about 10 pb$^{-1}$ 
collected at $E_b=935$ and $936$ MeV in 2019 and 2020).
The cosmic-ray distribution $H_{\rm csm}$ is measured with a special
cosmic-ray selection: $E_{\rm EMC}>0.7$ GeV,
a cosmic-ray track, and a signal in the muon system.  

The fit results is shown in the Fig.~\ref{fig:his19}. It is seen that the 
function (\ref{alen}) reproduces the shape of the $n\bar{n}$ distribution
reasonably well.

The fitted numbers of $n\bar{n}$ events for 7 energy points of the
2019 run are listed in Table~\ref{tab:crsect}. The quoted errors are
statistical and systematic. The sources of the systematic uncertainty are the
uncertainty in $\sigma_{\rm G}$, uncertainty in the time shift between energy
points above the $n\bar{n}$ threshold and below it, where 
$H_{\rm bkg}$ is determined, statistical fluctuations in $H_{\rm
bkg}$, and dependence of $H_{\rm csm}$ on selection criteria. The time shift
measured using $e^+e^-\to \gamma\gamma$ events varies from 
$-0.15$ ns to 0.25 ns. We conservatively estimate that the uncertainty
in the shifts does not exceed 0.1 ns. The uncertainty due to the statistical
fluctuations of $H_{\rm bkg}$ is estimated using toy MC study.
The total systematic uncertainty listed in Table~\ref{tab:crsect} grows with
energy due to increasing overlap of signal and background distributions.

In total, about 1250 $n\bar{n}$ events are selected in the 2019 data set.
The effective cross section for the beam-induced and physical background 
$\sigma_{\rm bkg}=N_{\rm bkg}/L$, where $L$ is the integrated luminosity for
a given energy point, is found to be independent of energy within the
statistical errors. Its average value over 7 energy points $5.1\pm1.1$ pb
is consistent with the value $3.9\pm1.0$ pb measured below the $n\bar{n}$
threshold.  The contribution to $\sigma_{\rm bkg}$
from the physical background is estimated using MC simulation. It is
dominated by the processes $e^+e^-\to K_SK_L\pi^0$,
$K_SK_L 2\pi^0$, and $K_SK_L\eta$, and is comparable to the value
obtained from the fit to data. 

The parameter $\alpha_n$ does not have a clear energy dependence. It
varies from 0.02 to 0.07 with a statistical error of about 0.01. 
Its average value is $\alpha_n=0.037\pm 0.07$ cm$^{-1}$ may be the
result of incorrect simulation of antineutron scattering and the fraction of
events rejected by the condition $E_3<0.75E_b$, while a large nonstatistical
spread arises presumably from uncertainties and shifts in time calibration.

\section{Determining the number of $n\bar{n}$ events for 
the 2017 data set \label{sec:Tim17}}
%============================= Fig. N ================================
\begin{figure*}
\includegraphics[width=0.47\textwidth]{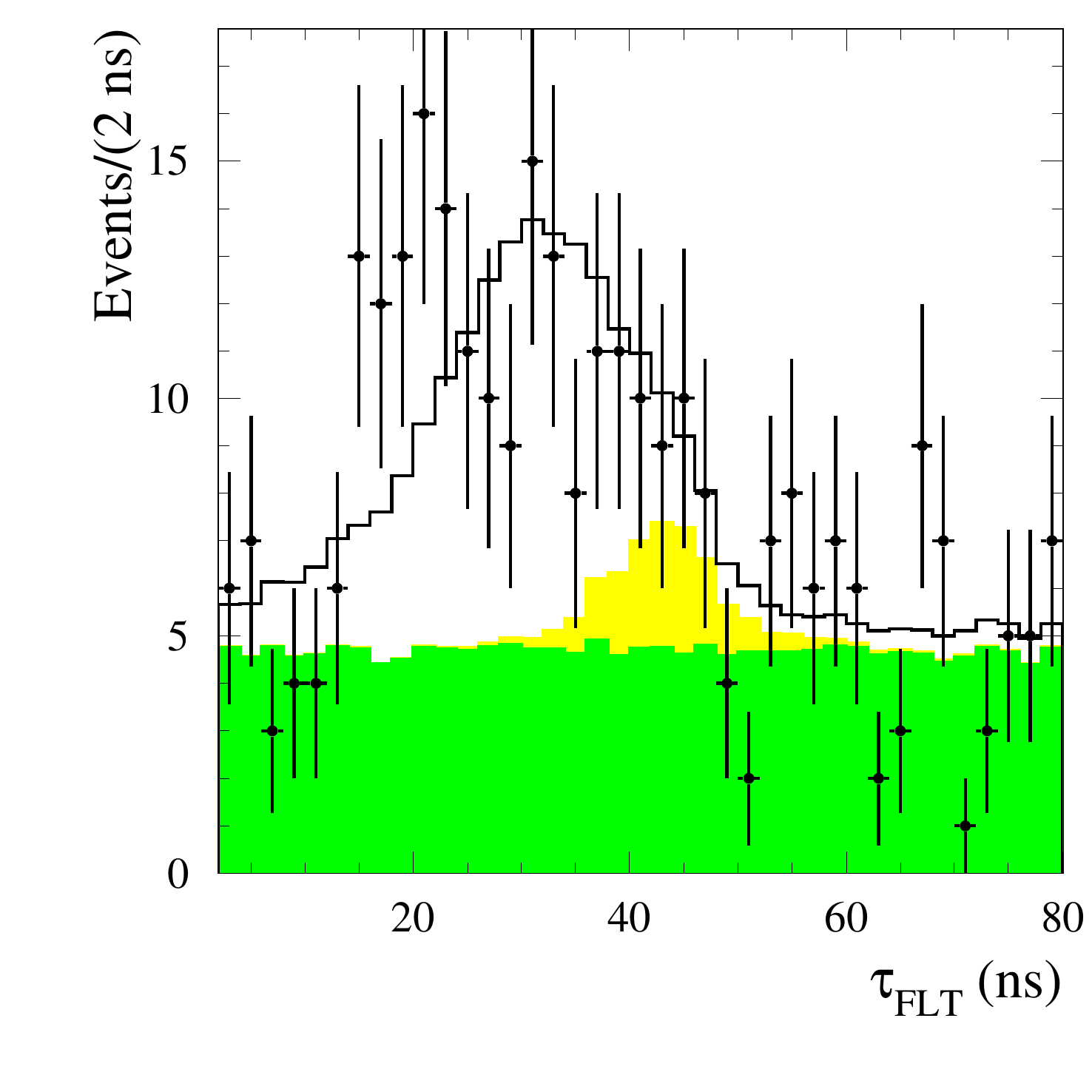} \hfill
\includegraphics[width=0.47\textwidth]{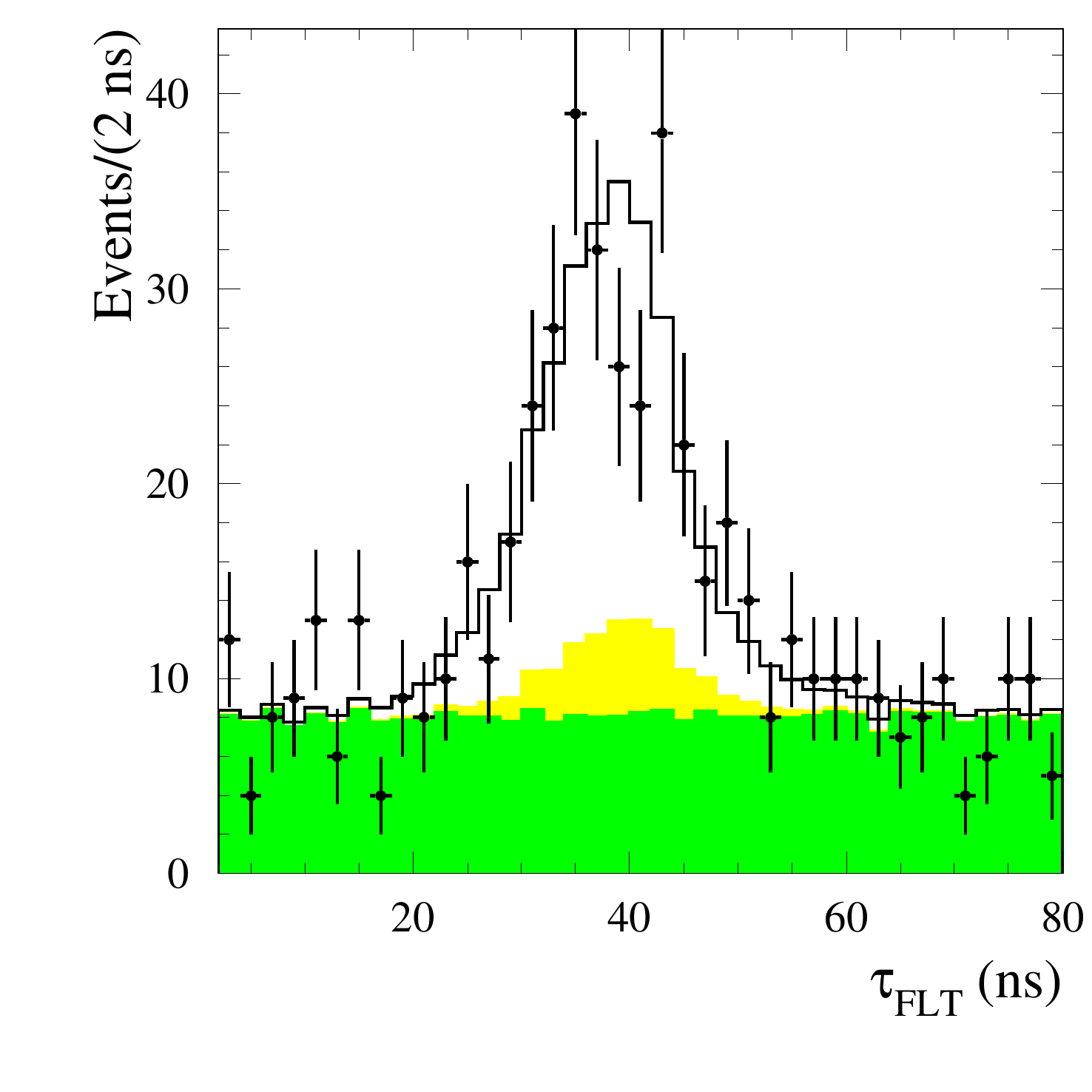} 
\caption {The $\tau_{\rm FLT}$ distribution
for selected data events collected in 2017 (points with error bars) at 
$E_b=942$ MeV (left panel) and at $E_b=1003$ MeV (right panel). 
The solid histogram is the result of the fit described in the text. 
The light-shaded (yellow) histogram shows the fitted cosmic-ray background.
The medium-shaded (green) region represents
the fitted beam-induced plus physical background.
\label{fig:his17}}
\end{figure*}
In analysis of the 2017 data set, we measure the time difference 
$\tau_{\rm FLT}$ between the signal of the EMC first level trigger 
(FLT)~\cite{SNDet1} and 
the beam revolution frequency with a rather poor resolution, about 8 ns for
$e^+e^-\to\gamma\gamma$ events. Such a time
resolution does not allow to separate $n\bar{n}$ events from the physical
and beam-induced backgrounds, but is sufficient to
measure and subtract the cosmic-ray background. 

The data $\tau_{\rm FLT}$ distributions for two energy points are
shown in Fig.~\ref{fig:his17}. Note that the time axis is reversed
so that delayed evens are located on the left side of the plot.
The distributions are fitted by Eq.~(\ref{eq17}) with the parameters 
$N_{n\bar{n}}$ and $N_{\rm csm}$ floating. 

Our Monte Carlo (MC) simulation does not include simulation of the time
distribution for the FLT signal. The $\tau_{\rm FLT}$ resolution function can
be obtained using data $e^+e^-\to\gamma\gamma$ events. However, the shape of 
this function depends on the distribution of the energy deposition in an event
over the calorimeter crystals and is different for $n\bar{n}$ and
$\gamma\gamma$ events. 
This difference is studied on the 2019 data set, where both methods of
time measurement can be used. From analysis of the
$\tau_{\rm FLT}+\tau_{\rm EMC}$ distributions for $n\bar{n}$ and
$\gamma\gamma$ events, we extract the time shift $\Delta t=2.4\pm0.1$
ns and the standard deviation $\sigma_G=3.7\pm0.5$ ns of the Gaussian
function, which is used to smear the $\gamma\gamma$ resolution function.

The signal distribution $H_{n\bar{n}}$ is obtained by convolution
the time spectrum of antineutron annihilations extracted from simulation
with the resolution function. The simulated events are previously reweighed
to take into account difference between data and simulation in the 
$\tau_{\rm EMC}$ spectrum observed in Sec.~\ref{sec:Tim19}. In addition to 
the procedure described in Sec.~\ref{sec:Tim19}, Eq.~(\ref{alen}) with 
$\alpha_n=0.037\pm 0.07$ cm$^{-1}$ is used for reweighing.
The  cosmic distribution $H_{\rm csm}$ is measured as described in
Sec.~\ref{sec:Tim19}. 

From the analysis of the 2019 data set (see Sec.~\ref{sec:Tim19}) we
find that $N_{\rm bkg}\simeq L\sigma_{\rm bkg}$, and that $\sigma_{\rm bkg}$
is weakly dependent on energy. The cross section $\sigma_{\rm bkg}$ and the
shape of the background distribution are measured
using data with an integrated luminosity of about 10 pb$^{-1}$ collected in
2017 below the $n\bar{n}$ threshold ($E_b=930$--938 MeV). The shape 
$H_{\rm bkg}$ is described reasonably well by the $\tau_{\rm FLT}$
distribution for data $e^+e^-\to\gamma\gamma$ events. The fitted background
cross section $\sigma_{\rm bkg}=12\pm3$ pb is significantly larger than the 
value $5.1\pm1.1$ pb obtained for the 2019 data set. We study predominantly 
background events with $0.8E_b<E_{\rm EMC}<0.9E_b$ and find that the
beam-induced background in 2017 is 3-4 times greater than in 2019.
Therefore, we conclude that the beam-induced background dominates in 
$\sigma_{\rm bkg}$ in 2017. The difference in the effective cross section
for background events with $0.8E_b<E_{\rm EMC}<0.9E_b$ between energy 
points above and below the $n\bar{n}$  threshold reaches 40\%.
This value is taken as an estimate of the systematic uncertainty in
$\sigma_{\rm bkg}$ for the standard selection.

The results of the fit is demonstrated in Fig.~\ref{fig:his17}.
The obtained numbers of $n\bar{n}$ events for 7 points of the 2017 run 
are listed in Table~\ref{tab:crsect}. The quoted errors are
statistical and systematic. The sources of the systematic uncertainty are the
uncertainties in the parameters $\sigma_{\rm G}$, $\Delta t$, $\alpha_n$, and
$\sigma_{\rm bkg}$. The uncertainty of $\sigma_{\rm bkg}$ gives dominant 
contribution. 

%============================= Fig. N ================================
\begin{figure*}
\includegraphics[width=0.32\textwidth]{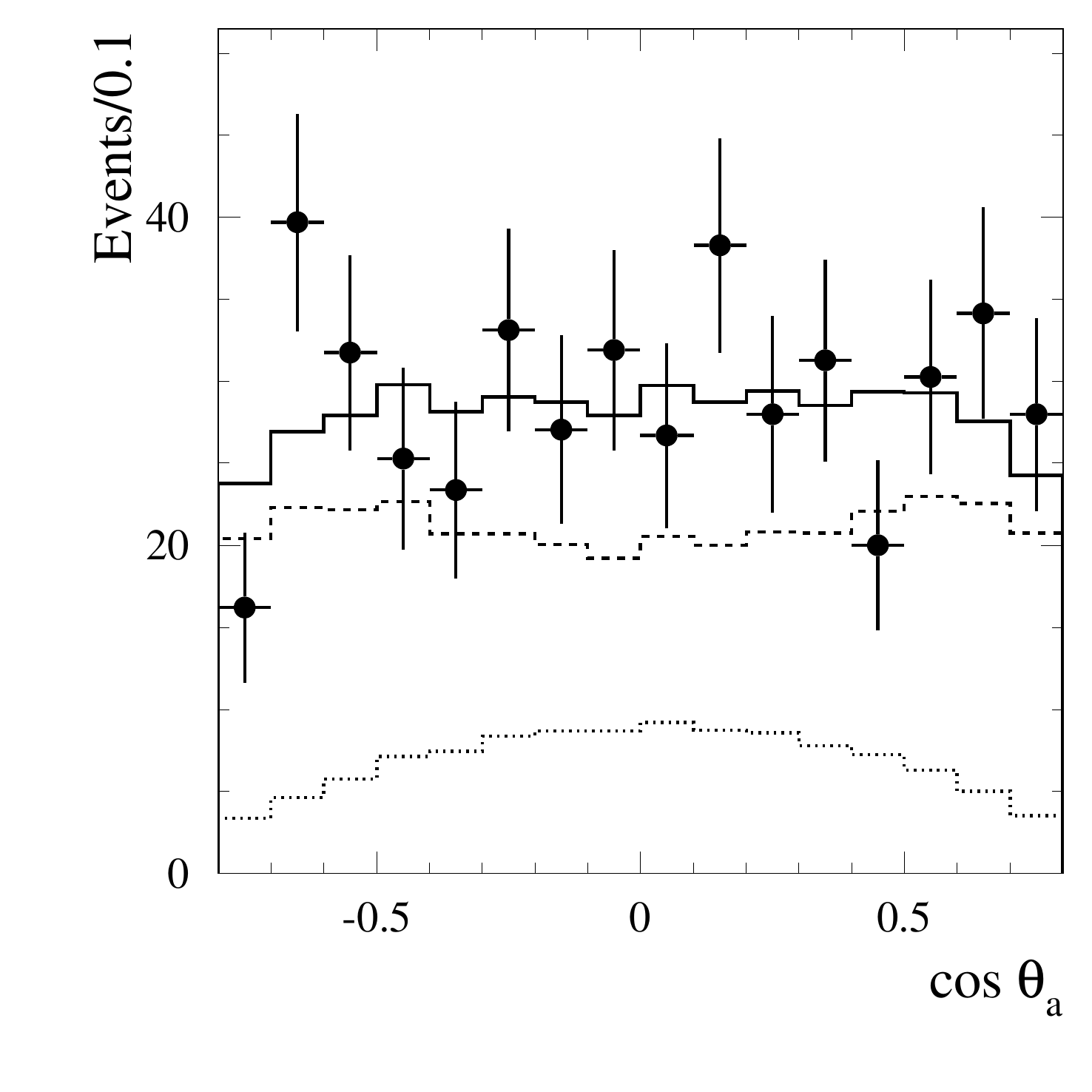} 
\includegraphics[width=0.32\textwidth]{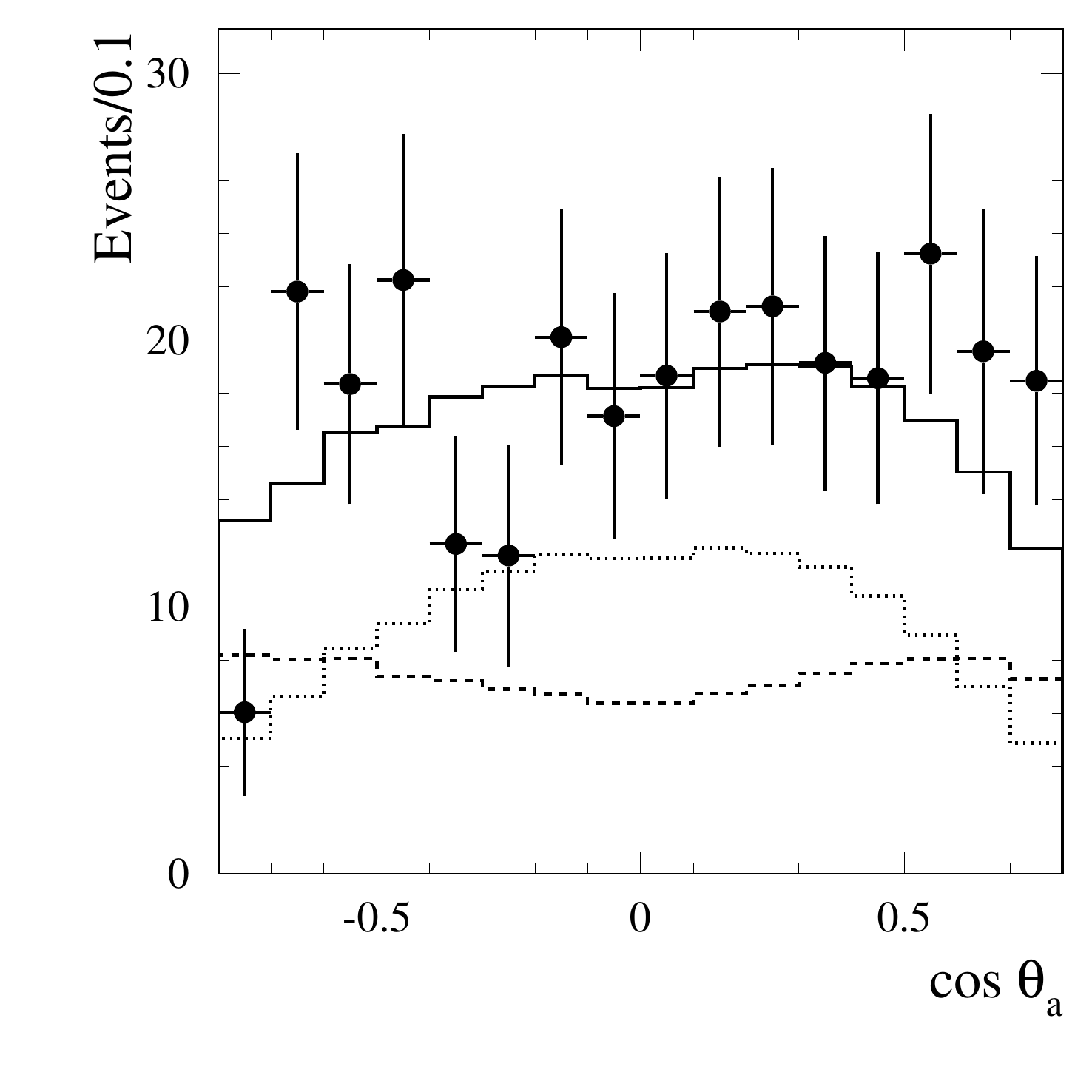}
\includegraphics[width=0.32\textwidth]{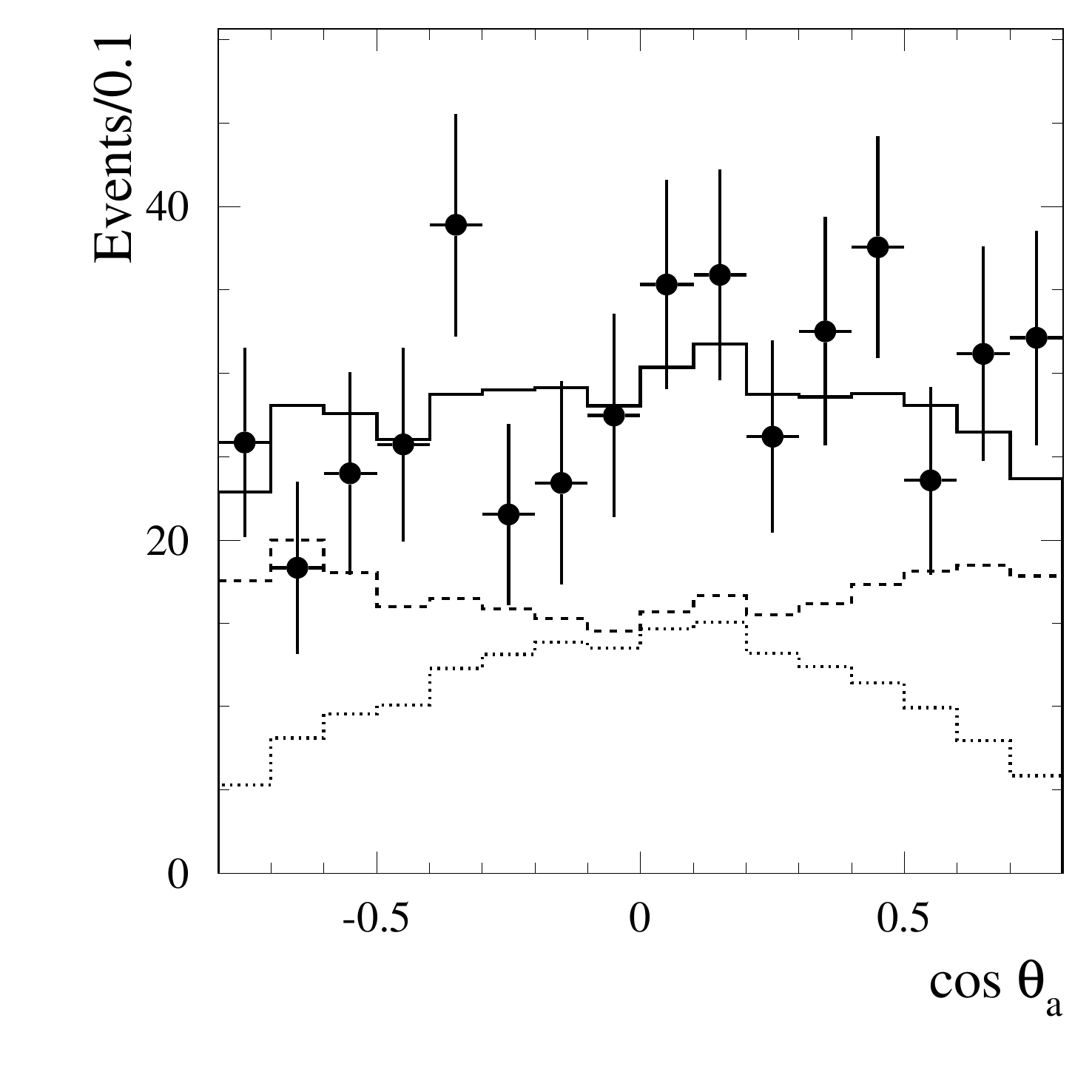}
\caption {The $\cos\theta_a$ distributions for data $n\bar{n}$ events 
of the the 2019 run (points with error bars) with $E_b=945$, 950, and 951 MeV
(left panel), with $E_b=956$, and 963 MeV (middle panel), and with
$E_b=973$, and 988 MeV (right panel).
The solid histogram is the result of the fit described in the text.
The dashed and dotted histograms shows the fitted contributions for the
magnetic and electric form factors, respectively.\label{fig:gemp2}}
\end{figure*}
%
%============================= Fig. N ================================
\begin{figure*}
\includegraphics[width=0.47\textwidth]{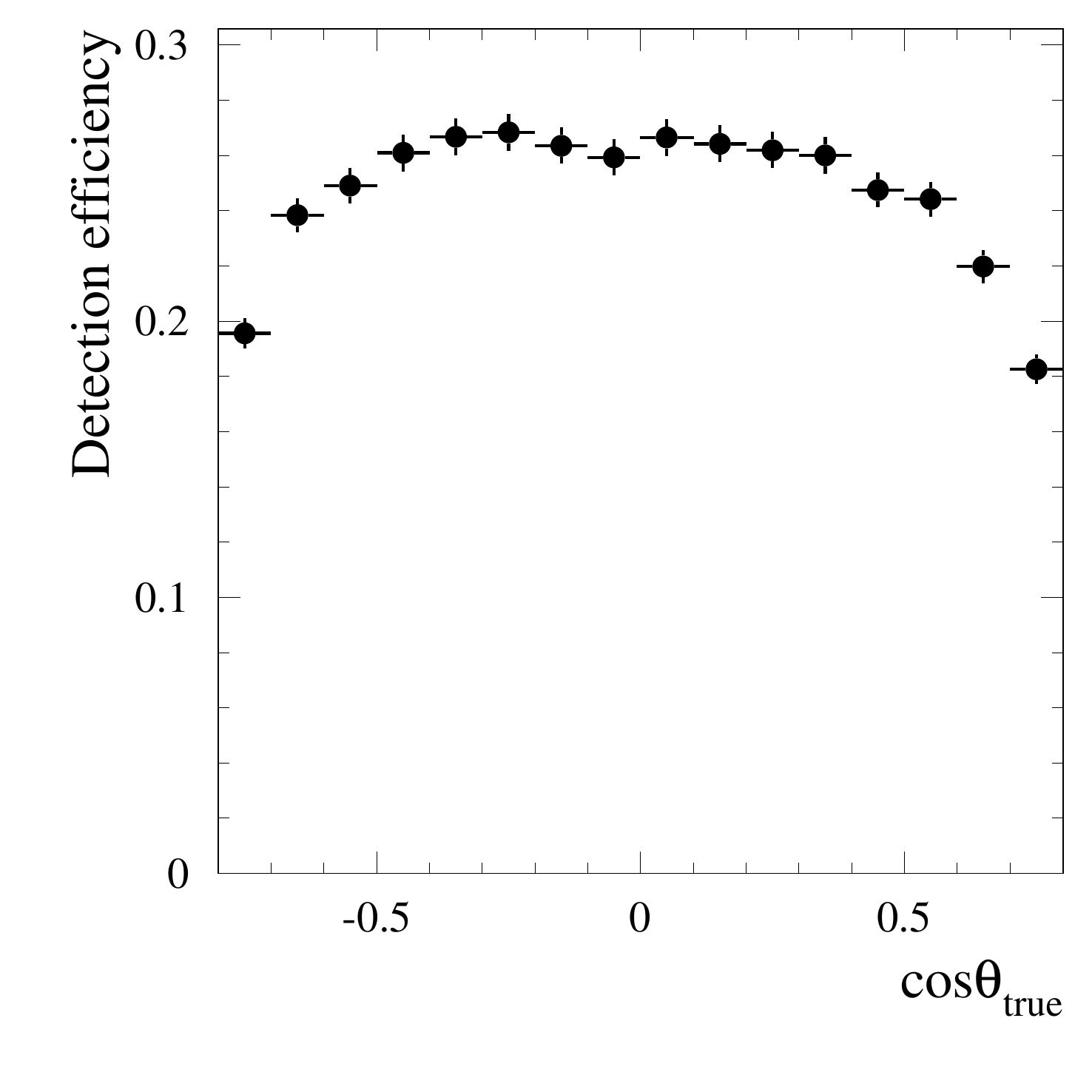} \hfill
\includegraphics[width=0.47\textwidth]{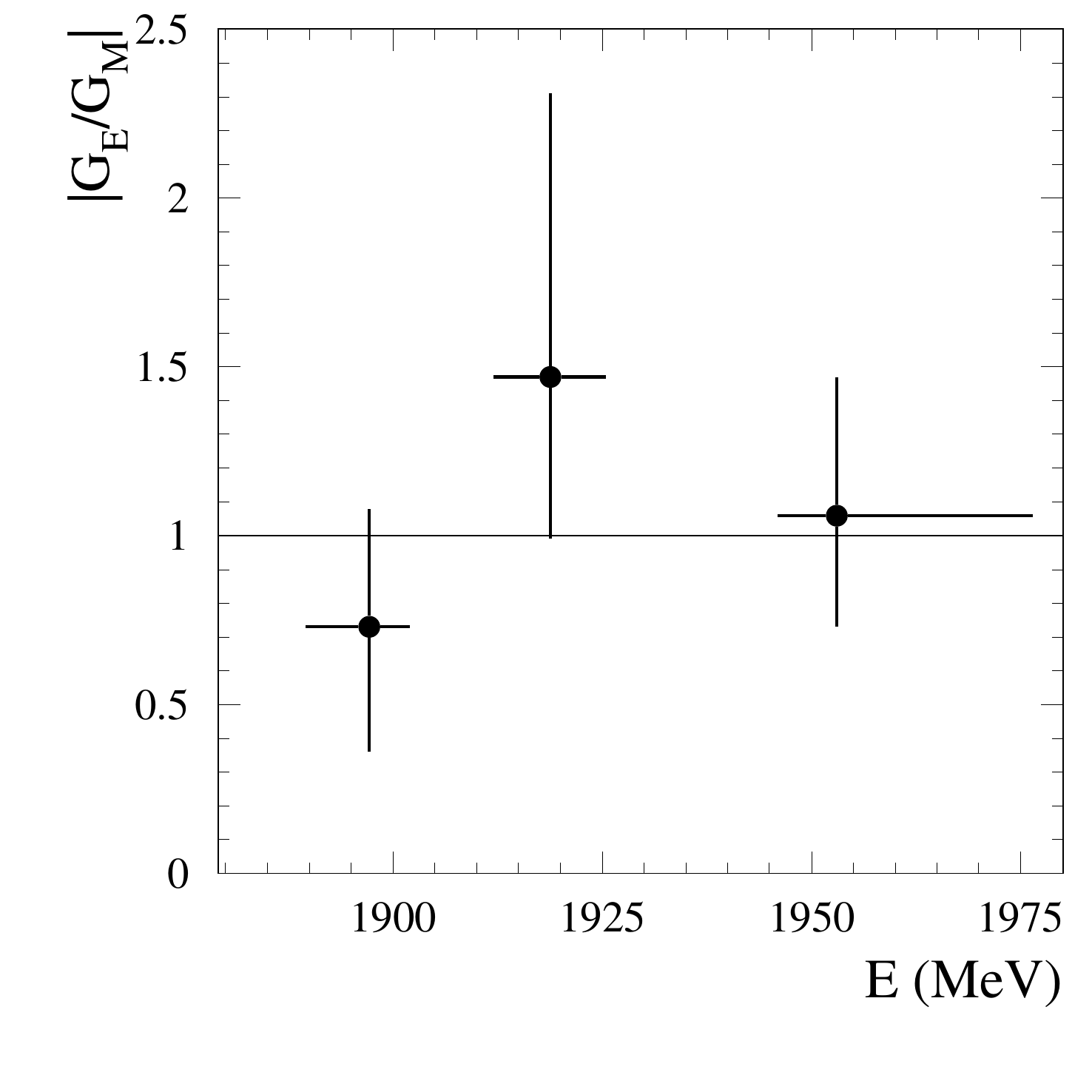} \\ 
\parbox[h]{0.47\textwidth}{\caption {The detection efficiency
determined using MC simulation as a function of $\cos{\theta_{\rm
true}}$, where $\theta_{\rm true}$ is the generated antineutron polar
angle.}
\label{fig:efth}} \hfill
\parbox[h]{0.47\textwidth}{\caption{The neutron $|G_E/G_M|$ ratio 
measured using the 2019 data set. The left edge of the plot
corresponds to the $n\bar{n}$ threshold, where $|G_E/G_M|=1$.}
\label{fig:gemn1}}
\end{figure*}
\section{Analysis of the antineutron angular distribution\label{sec:GeGm}}
%============================= Table N ================================
\begin{table}
\centering
\caption{The measured $|G_E/G_M|$ values.
\label{tab:gegm}}
\begin{tabular}{cc}
\hline
$E_b$ (MeV) & $|G_E/G_M|$ \\
945, 950, 951 &  $0.73^{+0.35}_{-0.37}\pm0.03$ \\ 
956, 963 & $1.47^{+0.84}_{-0.48}\pm0.12$ \\
973, 988 & $1.06^{+0.41}_{-0.33}\pm0.09$ \\
\hline 
\end{tabular}
\end{table}
The 2019 data set is used for analysis of angular distributions. For each
energy point, the range $-0.8< \cos\theta_a <0.8$ is divided into 16
intervals. Then, in each $\cos\theta_a$ interval, a fit is performed to the
$\tau_{\rm EMC}$ distribution as described in Sec.~\ref{sec:Tim19} (but with
$\alpha_n$ fixed at its average value). The obtained seven $\cos\theta_a$ 
distributions are combined into 3 distributions for the following groups 
of energy points: (945, 950, 951), (956, 963), (973,988), where the numbers
in parenthesis represent the values of $E_b$ in MeV. These distributions are
shown in Fig.~\ref{fig:gemp2}. They are fitted with the function
\begin{eqnarray}
F(\cos\theta_a)&=&A\bigg( H_M(\cos\theta_a)\nonumber\\
&+&\frac{1}{\gamma^2}\left |\frac{G_E}{G_M} \right |^2 H_E(\cos\theta_a)
\bigg),\label{eqff}
\end{eqnarray}
where $H_M$ and $H_E$ are the $\cos\theta_a$ distributions for selected 
simulated $n\bar{n}$ events generated with the angular distributions 
$1+\cos^2\theta$ and $\sin^2\theta$ (see Eq.~(\ref{eqB1})), respectively,
and $A$ is a normalization factor.
The shape $H_M$ and $H_E$ distributions differ from the generated initial
distributions due to nonuniform detection efficiency 
(see Fig.~\ref{fig:efth}) and the finite $\theta_a$ resolution, which 
has $\sigma=8^\circ$. 

The results of the fit are shown in Fig.~\ref{fig:gemp2}. The fitted 
$|G_E/G_M|$ values for three energy groups are listed in Table~\ref{tab:gegm}.
To estimate the systematic uncertainty, we vary the parameters $\alpha_n$ and 
$\sigma_G$ used in the fit to the $\tau_{\rm EMC}$ distributions within their 
uncertainties, introduce the $\tau_{\rm EMC}$ shift ($\pm0.1$ ns), and modify 
the background shape as described in Sec.~\ref{sec:Tim19}. From MC
simulation we find that most of the used selection criteria  do not have a
significant effect on the shape of the antineutron angular distribution.
The exceptions are the conditions $n_{\rm ch}=0$ and $L_\gamma>-2.5$. 
We exclude the condition $L_\gamma>-2.5$ (this leads to a tenfold
increase in the physical background), determine the $|G_E/G_M|$ ratios,
and take the difference between the the values obtained with different
selection criteria as an estimate of the systematic uncertainty. To
test the effect of the condition $n_{\rm ch}=0$, we add events containing 
one or several off-center charged tracks (see Sec.~\ref{sec:Efficy}), and 
again study a shift in the $|G_E/G_M|$ value. The systematic uncertainties 
from all sources are combined in quadrature.

Our results agree with the assumption that $|G_E/G_M|$=1, but 
also do not contradict larger values $|G_E/G_M|\approx 1.4$--1.5 observed
in the BABAR~\cite{Babar} and BESIII~\cite{BESpp} experiments for the
ratio of the proton form factors near $E=2$ GeV.
\section{Detection efficiency\label{sec:Efficy}}
%============================= Fig. N ================================
\begin{figure}
\centering
\includegraphics[width=0.45\textwidth]{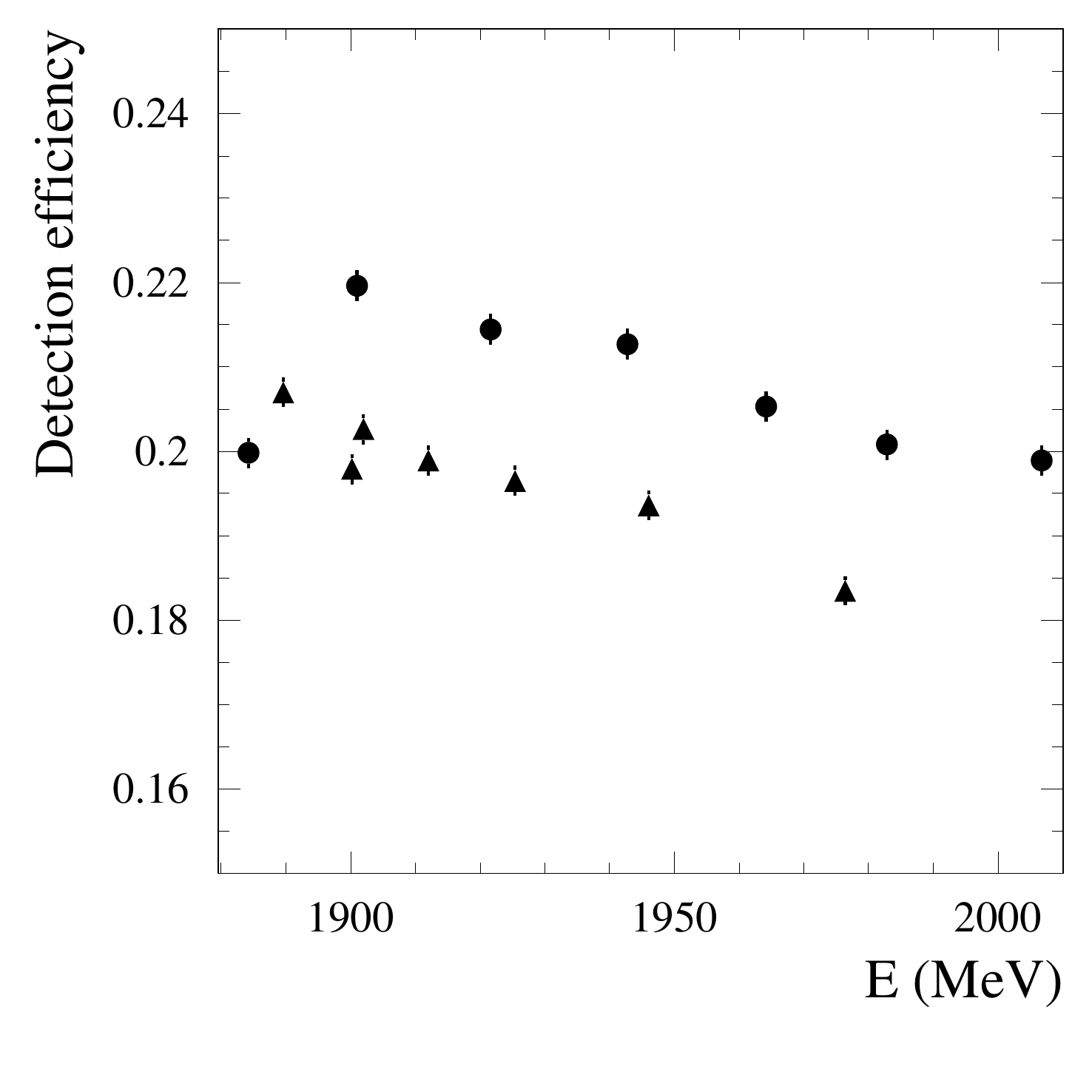}
\caption {  The energy dependence of the detection efficiency for
$e^+e^-\to n\bar{n}$ events determined using MC simulation for the 
2017 (circles) and 2019 (triangles) runs. 
The left edge of the plot corresponds to the $n\bar{n}$ threshold. 
\label{fig:efenr}}
\end{figure}
At first approximation, the detection efficiency $\varepsilon$ is calculated
using MC simulation with an angular distribution corresponding $|G_E/G_M|=1$.
The simulation includes 
the c.m. energy spread, which is about 1 MeV, and 
emission of an additional photon by initial electron and positron. It also
takes into account spurious beam-generated photons and charged
tracks. They are simulated by using special background events recorded during
data taking with a random trigger. These events are superimposed on the 
simulated $n\bar{n}$ events. The detector response is simulated with the 
GEANT4 toolkit~\cite{GEANT4}, release 10.5. The energy dependence of the
detection efficiency obtained with our standard selection criteria (see 
Sec.~\ref{sec:EvSelect}) is shown in Fig.~\ref{fig:efenr}. 

The decrease in the efficiency when approaching to the $n\bar{n}$ threshold
is due to an increase of the fraction of antineutrons annihilating before
calorimeter and producing charged tracks. The decrease of the
efficiency with increasing energy is due to an increase of the
probability of antineutron passing through the calorimeter without
interaction. 

The detection efficiency for the 2017 run is about
10\% higher than for 2019. The reason is the difference in the calorimeter 
digitizing electronics used in these runs. This leads, in particular, to 
a larger numbers of fired crystals with low amplitudes in 2017. 
Therefore, the $E_{\rm EMC}$ and $R_T$ (see Sec.~\ref{sec:EvSelect}) 
distributions for the 2017 and 2019 data sets are different. Of the 10\% 
difference in efficiency, 3\% and 6\% are due to the conditions on the 
parameters $E_{\rm EMC}$ and $R_T$, respectively.

As shown in the previous section, the measured ratio $|G_E/G_M|$ agrees with
unity in the energy region under study. To take into account its possible 
deviation from unity and the associated change in the antineutron angular
distribution, we introduce a model uncertainty in the detection efficiency of
6\%. This value corresponds the $|G_E/G_M|$ variation from 0.4 to 1.7.
%
%============================= Fig. N ================================
\begin{figure*}
\includegraphics[width=0.32\textwidth]{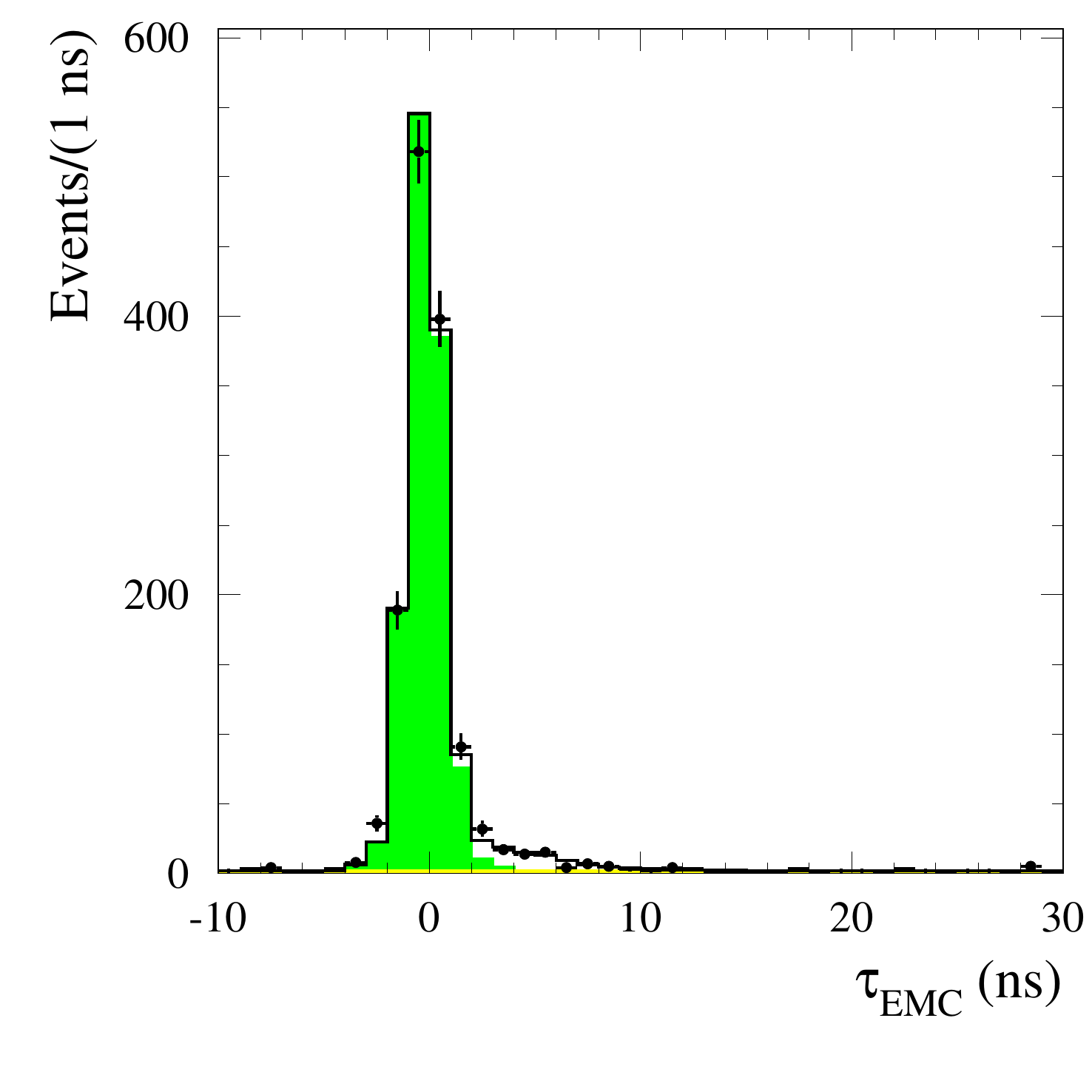} 
\includegraphics[width=0.32\textwidth]{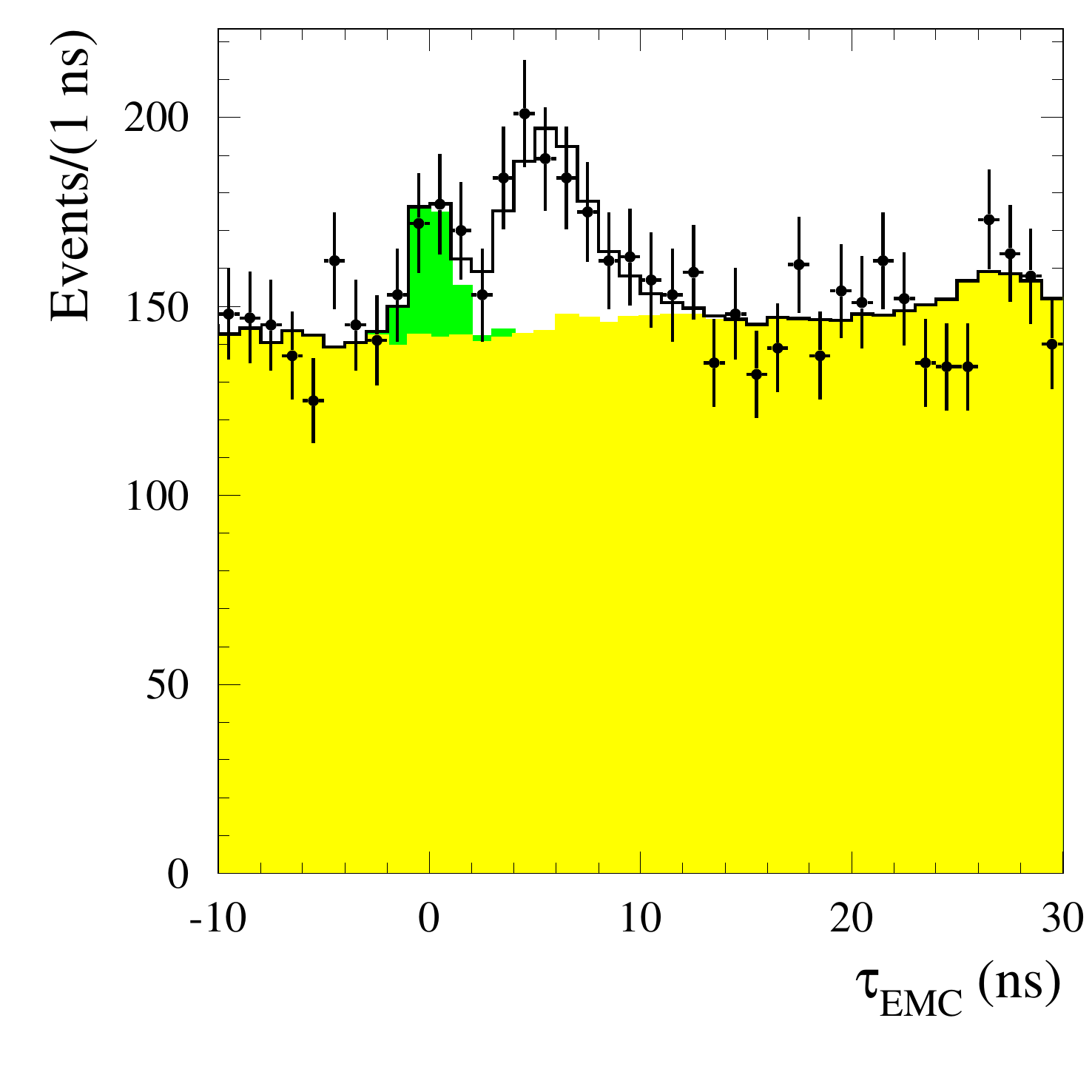}\hfill
\includegraphics[width=0.32\textwidth]{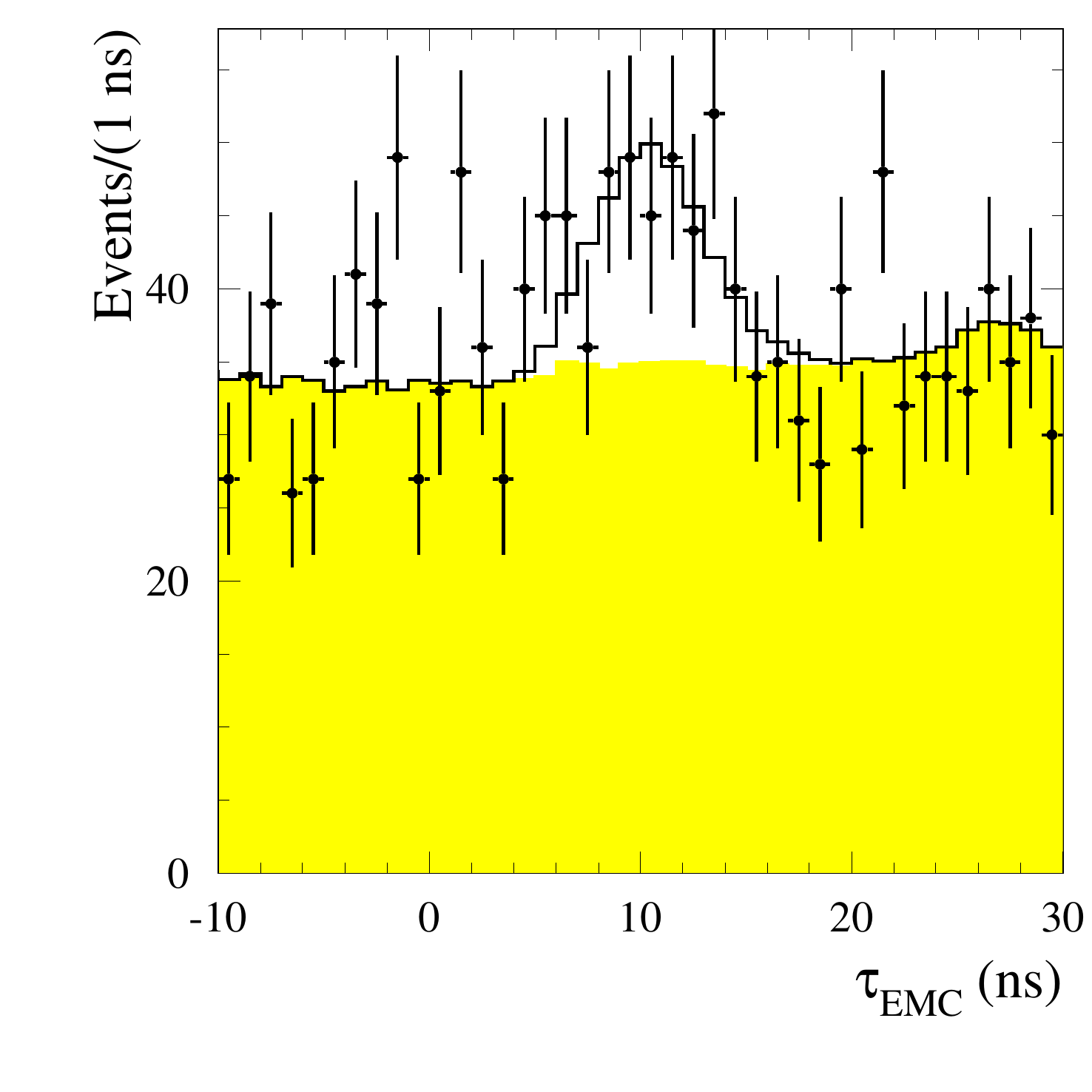} 
\caption { The $\tau_{\rm EMC}$ distributions for data events with
$E_b=950$, 951, 956 MeV from the 2019 data set selected with inverted
conditions on the parameters $5\land 6\land 7\land 8$ (left), 9 (middle), and
10 (right), where the condition numbers from Sec.~\ref{sec:EvSelect} are used.
The solid histogram is the result of the fit described in the text. 
The light-shaded (yellow) histograms show the fitted cosmic-ray background. 
The medium-shaded (green) regions at the left and middle plots represent the
fitted beam-induced plus physical background.
\label{fig:testsel}}
\end{figure*}

The detection efficiency is corrected for the difference in detector
response for $n\bar{n}$ events between data and MC simulation. 
The number of $n\bar{n}$ events for the 2019 data set can be determined
using significantly looser selection criteria than the standard ones.
We invert one of the selection conditions described in
Sec.~\ref{sec:EvSelect} and calculate the efficiency correction for 
the difference between data and simulation associated with this condition 
as follows:
\begin{equation}
\delta_i=\frac{n_0}{n_0+n_1}\frac{m_0+m_1}{m_0}-1,
\label{eqr19}
\end{equation}
where $n_0$ and $n_1$ ($m_0$ and $m_1$) are the numbers of $n\bar{n}$
data (MC) events selected using the standard selection and the
selection with the inverted condition $i$, respectively. 
The number $n_1$ is determined from the fit to the $\tau_{\rm EMC}$
spectrum as described in Sec.~\ref{sec:Tim19}, but with $\alpha_n$ fixed
at its average value. The shape of the 
distribution for the beam-induced and physical backgrounds is found
using data recorded below the $n\bar{n}$ threshold. 

Examples of the $\tau_{\rm EMC}$ spectra obtained with inverted conditions 
$5\land 6\land 7$, 8, and 9 are shown in Fig.~\ref{fig:testsel}.
Here the condition numbers from Sec.~\ref{sec:EvSelect} are used.
It is interesting to note that the left (right) spectrum in
Fig.~\ref{fig:testsel} does not contain
the cosmic-ray (beam-induced + physical) background component. The
signal distribution in the right spectrum is delayed compared to the middle
spectrum because of the condition $E_3>0.75E_b$ selects events,
in which antineutrons annihilate predominantly in the third calorimeter layer.

The obtained corrections averaged over seven energy points are listed in 
Table~\ref{tab:effcor}. Condition 2 ($36^\circ<\theta_a<144^\circ$) is absent
in the table, since the model uncertainty associated with the antineutron 
angular distribution was considered above.
%============================= Table N ================================
\begin{table*}
\centering
\caption{The efficiency corrections for different conditions. The
condition numbers from Sec.~\ref{sec:EvSelect} are used.
\label{tab:effcor}}
\begin{tabular}{ccccccc}
\hline
Condition    & 1 & 3 & 4 & $5\land6\land 7\land 8$ & 9 & 10 \\
$\delta_i$, \% & $4.1\pm1.4$ & $0.9\pm1.4$ & $-9.8\pm1.9$ &$-2.3\pm1.5$ &
$9.4\pm3.9$ & $-5.3\pm2.2$ \\
\hline 
\end{tabular}
\end{table*}
%============================= Fig. N ================================
\begin{figure}
\begin{center}
\includegraphics[width=0.47\textwidth]{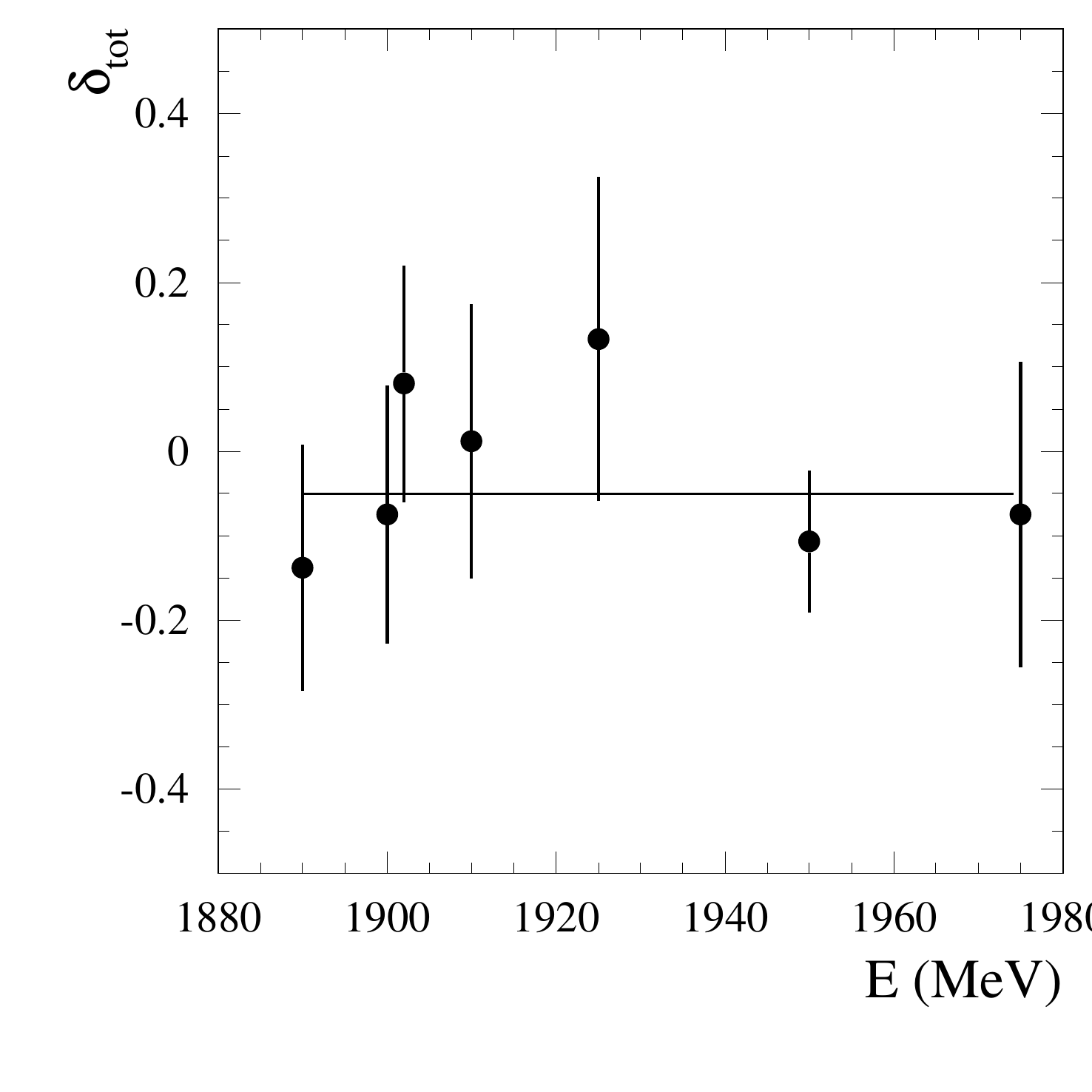}
\caption{The energy dependence
of the efficiency correction associated with the $n\bar{n}$ selection 
conditions. The line indicates a fit to a constant value.   
\label{fig:corvse}}
\end{center}
\end{figure}

For selection criterion 4, the condition $0.7E_b<E_{\rm EMC}<E_b$ is used
instead of full inversion. To determine the correction
associated with criterion 1 ($n_{\rm ch}=0$), we select events with one or 
several off-center charged tracks having $D_{xy}>0.5$ cm, where $D_{xy}$ is 
the distance between the track and the beam collision axis. The simulation
shows that about 20\% of $n\bar{n}$ events 
give tracks in the SND drift chamber, most of which are off-center.
At $E_b<960$ MeV antiprotons from the $e^+e^-\to p\bar{p}$ process annihilate
in the material before the drift chamber (at a radius of about 2 cm from the
beam axis) and produce events with topology similar to $n\bar{n}$ events.
To suppress the $p\bar{p}$ and beam-induced backgrounds, we additionally require
that the maximum over charged tracks $D_{xy}$ be greater than 2.3 cm. This 
condition leads to a loss about 30\% of $n\bar{n}$ events with charged tracks.
The remaining small $p\bar{p}$ background is subtracted using MC simulation.
It should be noted that a significant fraction of $n\bar{n}$ events with
charged tracks is rejected by the condition $L_\gamma>-2.5$. Therefore,
we remove this condition when determine $n_{0,1}$ and $m_{0,1}$ for the
correction associated with criterion 1. 

We do not observe significant dependences of the corrections on the
beam energy and, therefore, list in Table~\ref{tab:effcor} the values
averaged over seven energy points of the 2019 data set. 
The total correction is calculated as
\begin{equation}
\delta_{\rm tot}=\Pi(1+\delta_i)-1.
\label{eqr20}
\end{equation}
Its energy dependence shown in Fig.~\ref{fig:corvse} is well fitted by
a constant value of $(-0.050\pm0.051)$. This value is taken as an efficiency
correction for data-MC simulation difference in the selection conditions for
the 2019 data set.

%============================= Fig. N ================================
\begin{figure*}
\includegraphics[width=0.47\textwidth]{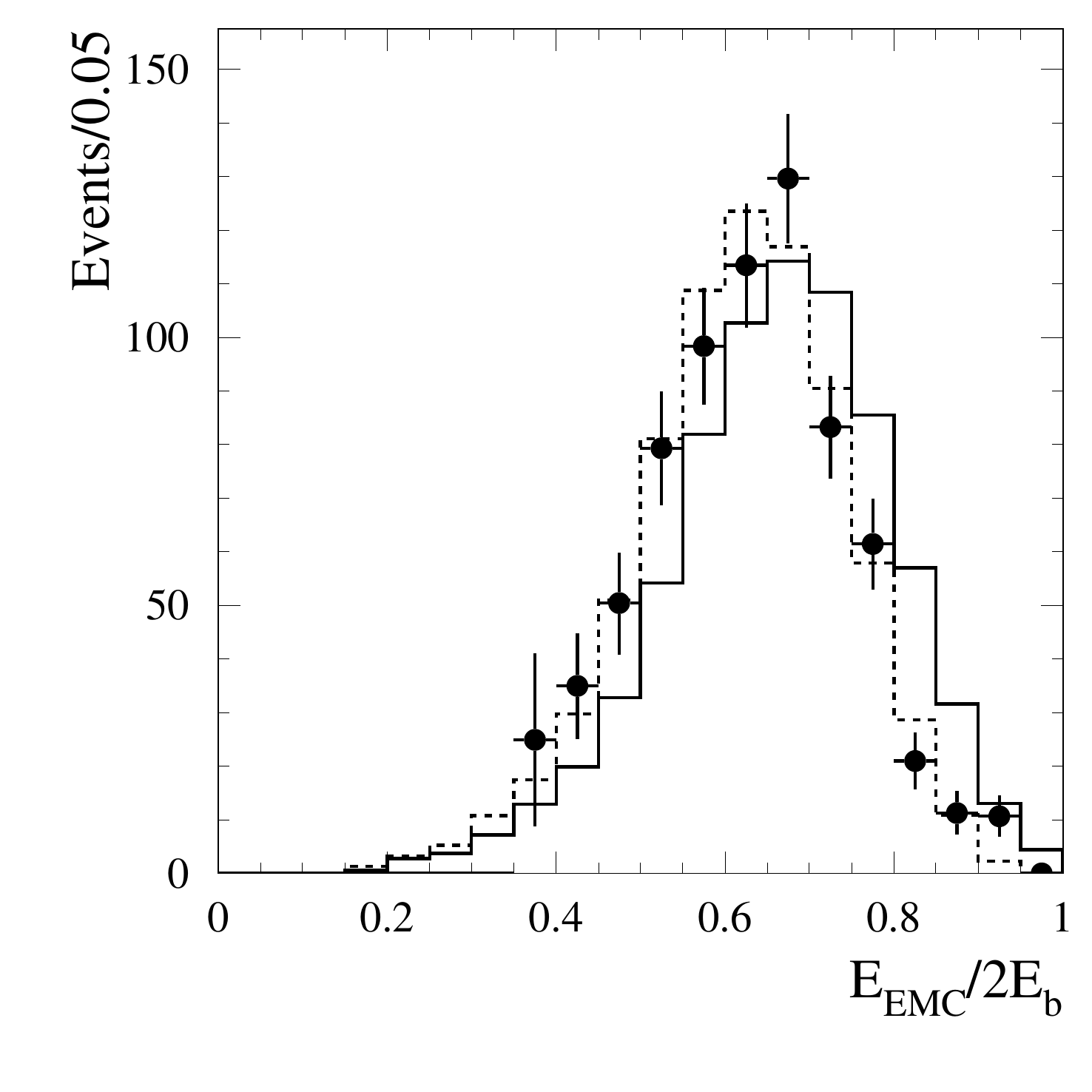} \hfill
\includegraphics[width=0.47\textwidth]{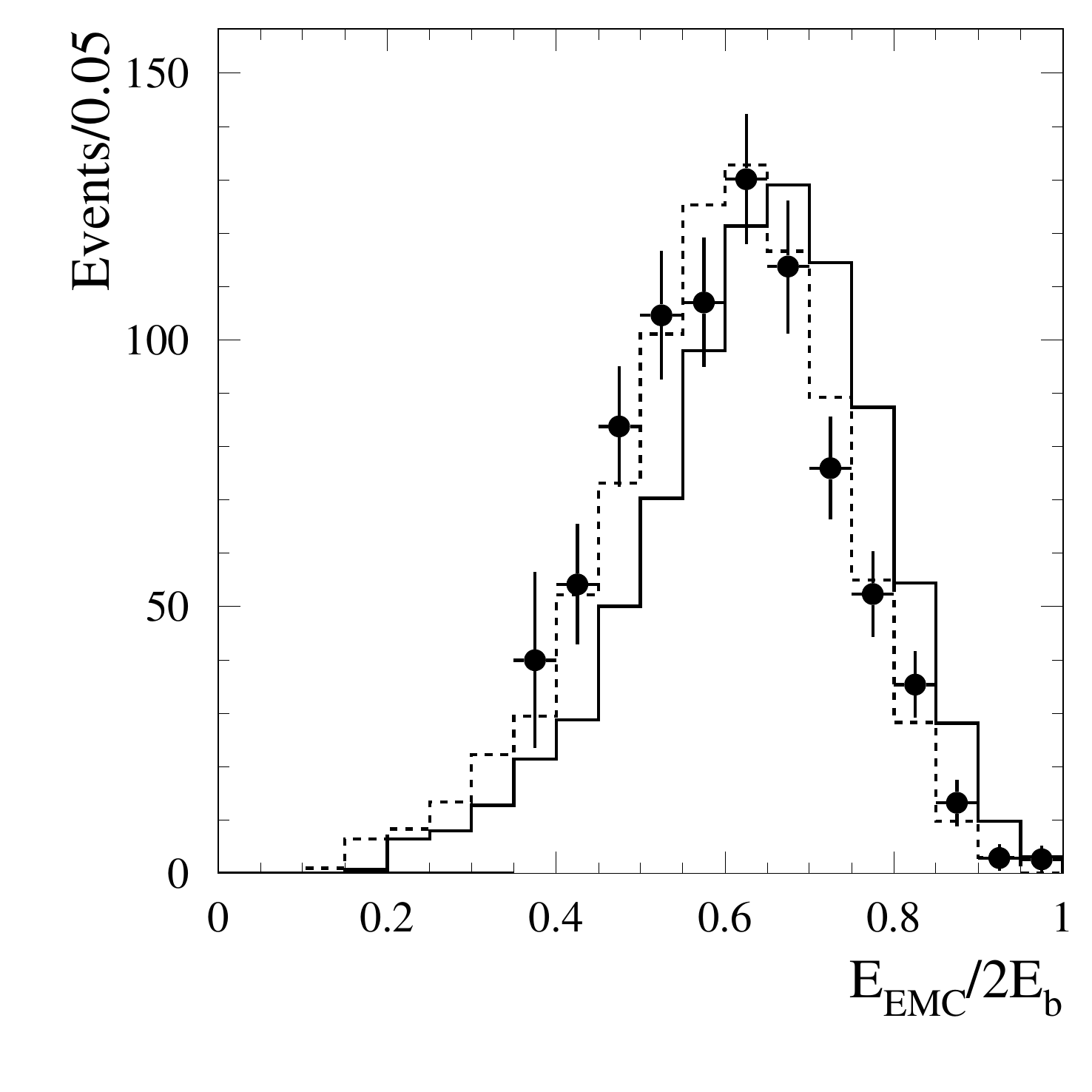} 
\caption {
The $E_{\rm EMC}/2E_b$ distribution for data $n\bar{n}$ events (points
with error bars) of the 2019 run with $E_b=945$, 950, 951, 956 MeV (left
panel), and with $E_b=963$, 973, 988 MeV (right panel). The solid 
histogram represents the same distribution for simulated $n\bar{n}$ events.
The standard selection criteria are used, except for the condition on 
$E_{\rm EMC}$. The dashed histograms are the simulated distributions after the
transformations $E_{\rm EMC} \to 0.93 E_{\rm EMC}$ at the left plot, and
$E_{\rm EMC} \to E_{\rm EMC}-100\mbox{ MeV}$ at the right plot.
\label{fig:emct}}
\end{figure*}

In the efficiency correction study above, the $E_{\rm EMC}$ threshold was
lowered to $0.7E_b$. To estimate systemic uncertainty associated with this
threshold, we compare the $E_{\rm EMC}/2E_b$ spectra 
for $n\bar{n}$ events in data and simulation. The spectra for two energy
intervals of the 2019 run are shown in Fig.~\ref{fig:emct}. The data spectra
are obtained by fitting the $\tau_{\rm EMC}$ distributions in each $E_{\rm
EMC}/2E_b$ bin as described in in Sec.~\ref{sec:GeGm}.

It is seen that the energy deposition in data is less than in simulation.
To match the data and MC spectra, we transform the simulation distribution
either by scaling $c E_{\rm EMC}$ or by shifting $E_{\rm EMC}-b$. Then 
the fraction of events rejected by the condition  $E_{\rm EMC}>0.7E_b$ is
recalculated. The result of scaling is shown in Fig.~\ref{fig:emct} (left) 
by the dotted histogram, while the result of shifting is presented in 
Fig.~\ref{fig:emct} (right). 
The fraction of events below the threshold $0.7E_b$ is about 3\% in
the range $E_b=945$--956 MeV and about 5\% in the range $E_b=963$--988 MeV.
The difference in this fraction after and before the shift transformation is 
taken as an estimate of the efficiency correction, while
the difference between the shift and scale transformations is taken as its
systematic uncertainty. The correction is found to be $(-2\pm1)\%$ below
956 MeV and  $(-3\pm2)\%$ above.

Some of the antineutrons pass through the calorimeter without interaction. 
Such events are not taken into account by the efficiency corrections described
above. In simulation their fraction increases from 0.5\% at $E_b=945$ MeV to
6.2\% $E_b=973$ MeV, and then to 9.4\% at $E_b=1003$ MeV.
In Sec.~\ref{sec:Tim19} we discuss the difference in the antineutron 
annihilation length between data and simulation and reweight simulated 
$n\bar{n}$ events to correct for this difference. With the reweighted simulation
the fraction of antineutrons passing through the calorimeter without 
interaction becomes 0.01\% at $E_b=945$ MeV, 3.2\% at $E_b=973$ MeV, and 5.5\% at
$E_b=1003$ MeV. The difference between the values obtained with unweighted
and weighted simulation with 100\% uncertainty is taken as an efficiency
correction.

For the 2017 data set, the efficiency corrections for all parameters except
$E_{\rm EMC}$ and $R_T$ are assumed to be the same as for 2019. The
$E_{\rm EMC}$ distributions for data and simulation at $E_{\rm EMC}>0.9E_b$ 
are compared between each other and with the same distributions for 2019.
For the parameter $R_T$, we loosen the condition on $R_T$ to $R_T>0.25$.
For the both parameters we do not observe deviations from the
corrections for the 2019 data set within statistical uncertainties.
Therefore, the same corrections are used for the both data sets, but for
2017 data a systematic uncertainty of 7\% associated with the parameters
$E_{\rm EMC}$ and $R_T$ is added.

The total efficiency correction for the 2019 run is $-(6.5\pm8.0)\%$ at 
$E_b=945$ MeV, $-(5.3\pm8.5)\%$ at $E_b=963$ MeV, and $-(4.6\pm8.7)\%$ at 
$E_b=988$ MeV. For the 2017 run, the total correction is
$-(6.9\pm10.6)\%$ at $E_b=942$ MeV, $-(5.1\pm11.1)\%$ at $E_b=971$ MeV, and
$-(4.3\pm11.4)\%$ at $E_b=1003$ MeV. The values of the corrected detection
efficiency and its systematic uncertainty are listed in in 
Table~\ref{tab:crsect}.
\section{The $e^+e^-\to n\bar{n}$ cross section and neutron effective
form factor \label{sec:Crosct}}
%============================= Fig. N ================================
\begin{figure}
\centering
\includegraphics[width=0.47\textwidth]{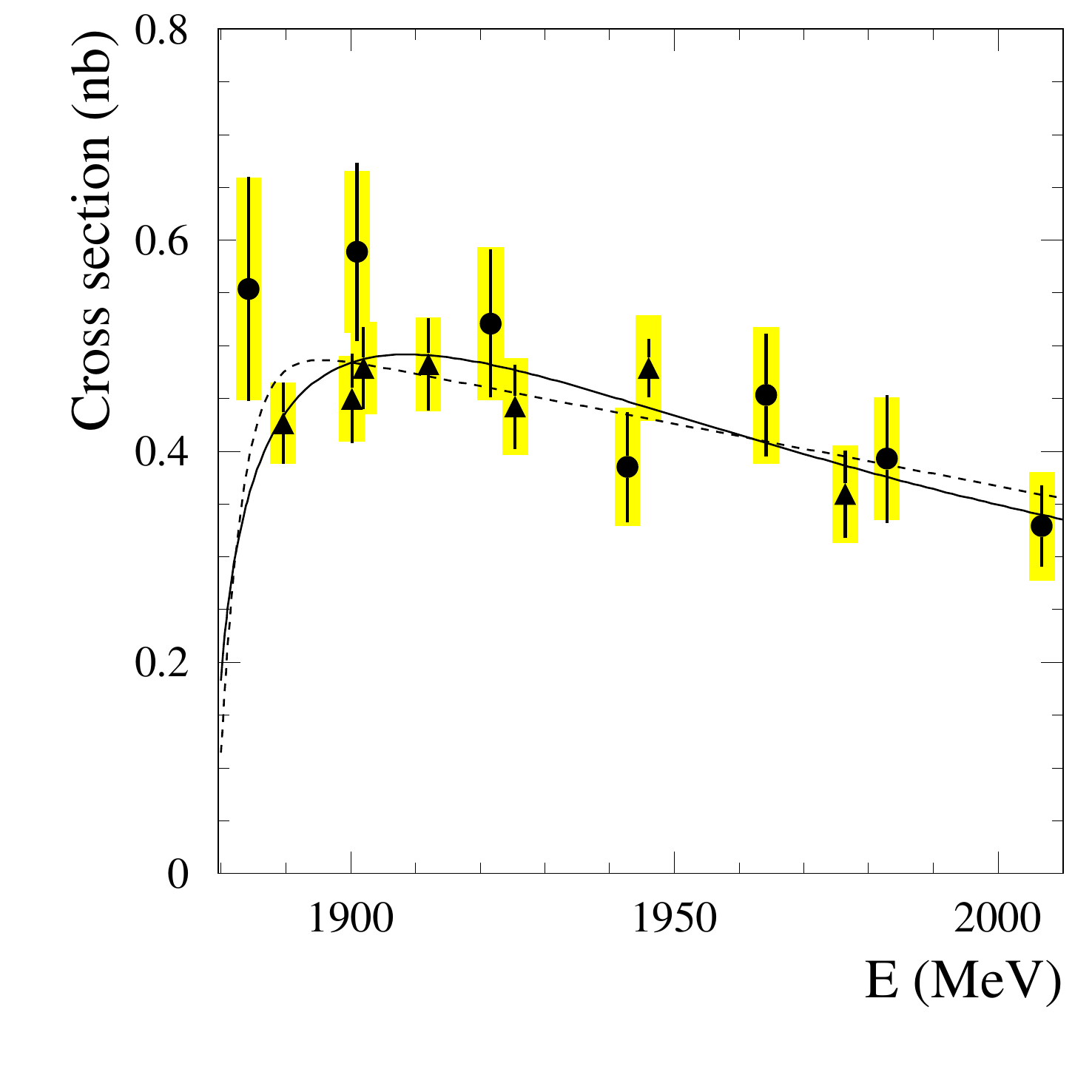}
\caption {The $e^+e^-\to n\bar{n}$ cross section measured
using the 2017 (circles) and 2019 (triangles) data sets. The error
bars and shaded boxes represent the statistical and systematic
uncertainties, respectively. The solid and dashed curves are the fit
results for Model I and II, respectively.
\label{fig:csect1}}
\end{figure}
\begin{figure}
\centering
\includegraphics[width=0.47\textwidth]{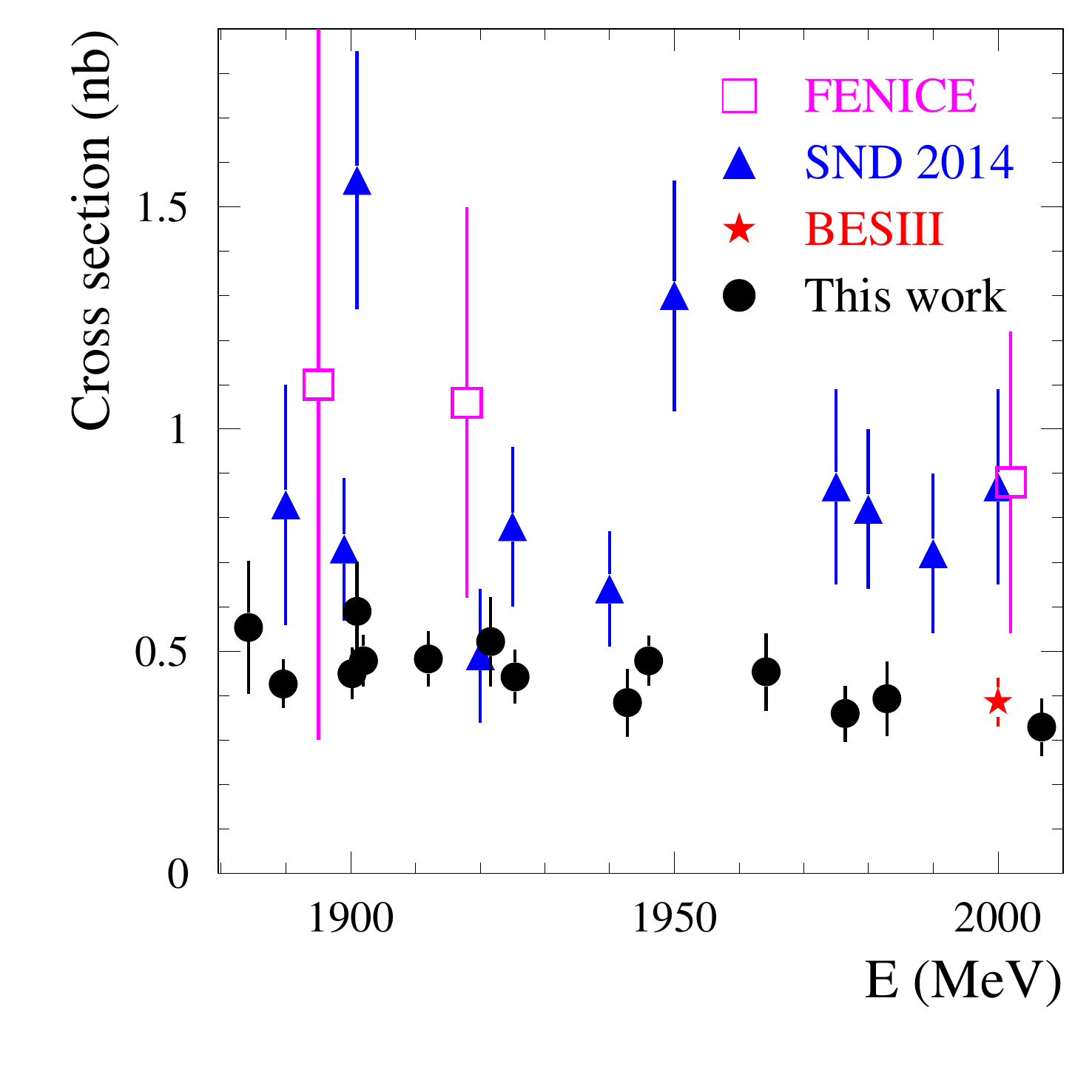}
\caption {The comparison 
of the $e^+e^-\to n\bar{n}$ cross section measured in this work
with the previous FENICE~\cite{FENICE}, SND~\cite{SNND}, and 
BESIII~\cite{BES} measurements. The combined statistical and
systematic uncertainties are shown for the new SND data.
\label{fig:csect2}}
\end{figure}
The visible cross section directly measured in experiment is related
to the Born cross section $\sigma$ as follows 
\begin{eqnarray}
\sigma_{vis}(E)&=&\sigma(E)(1+\delta(E))\nonumber\\
&=&\int_{-\infty}^{+\infty}G(E^\prime,E)dE^\prime\nonumber\\
&&\int_0^{x_{max}}W(s,x)\sigma(s(1-x))dx,
\label{eqB4}
\end{eqnarray}
where $G(E^\prime,E)$ is a Gaussian function describing the c. m. energy
spread, $W(s,x)$ is the radiator function~\cite{Radcor} describing emission
of photons by initial electrons and positrons, $x$ is a fraction of the beam
energy carried out by these photons, and $x_{max}=1-4m_n^2/s$. Here we
define the factor $(1+\delta)$, which takes into account the combined effect
of radiative corrections and beam energy spread. Equation~\ref{eqB4} is used
to fit the experimental data on the visible cross section 
$\sigma_{vis,i}=N_{n\bar{n},i}/(L_i\varepsilon_i)$, where $i$ is the index
of the energy point in Table~\ref{tab:crsect}. The Born cross section in the
fit is given by Eq.~\ref{eqB2}, where the effective form factor is 
parametrized by the second-order polynomial $|F|=a_0+a_1p_n+a_2p_n^2$
(Model I), where $p_n$ is the neutron momentum, and
$a_i$ are free fit parameters. After the fit, the factors $(1+\delta(E_i))$
are calculated using Eq.~\ref{eqB4}, and the experimental values of the Born 
cross section are obtained as $\sigma_i=\sigma_{vis,i}/(1+\delta(E_i))$.
To estimate uncertainties in $(1+\delta(E_i))$, we vary the parameters
$a_i$ within their errors and use the different 
parametrization for the Born cross section (Model II)
\begin{eqnarray}
\sigma(E)&=&b_1\left[1-\exp\left(-\frac{E-2m_n}{b_2}\right)\right]\nonumber\\
&&\left[1+b_3(E-2m_n)\right].
\end{eqnarray}
in the fit. Such parametrization is used to describe the energy dependence
of the $e^+e^-\to p\bar{p}$ cross section near the threshold in Ref.~\cite{cmd}. The 
difference in $(1+\delta(E_i))$ between Models I and II is
taken as an estimate of the model uncertainty. The total uncertainty in
$(1+\delta(E_i))$ is 2.2\% at $E_b=942$ MeV, 1.6\% at $E_b=1003$ MeV,
and does not exceed 1\% in other energy points.

The measured Born cross section is shown in Fig.~\ref{fig:csect1}
and listed in Table~\ref{tab:crsect}. The systematic uncertainty in the
cross section includes uncertainties in the number of $n\bar{n}$ events,
detection efficiency, factor $(1+\delta)$, and integrated luminosity.
The comparison of the cross section measured in this work
with the previous measurements is presented in Fig.~\ref{fig:csect2}.
Our cross section is about 0.4 nb and  considerably lower than 
the previous results of the FENICE~\cite{FENICE} and SND~\cite{SNND}
experiments. On the other hand, near $E=2$ GeV our result is in good agreement
with the BESIII measurement~\cite{BES}.

The previous SND results~\cite{SNND} are based on data collected in 2011 and 
2012. Reanalysis of the 2012 data set is performed using the selection
criteria and technique described in Secs.~\ref{sec:EvSelect} and 
\ref{sec:Tim17}, and MC simulation with GEANT4 version 10.5. Basing on this
reanalysis we conclude that the detection efficiency obtained from simulation
and the beam-induced background were underestimated in Ref.~\cite{SNND}.
The results on the $e^+e^-\to n\bar{n}$ cross section obtained in this
work supersede the measurements of Ref.~\cite{SNND}.
%============================= Fig. N ================================
\begin{figure}
\centering
\includegraphics[width=0.47\textwidth]{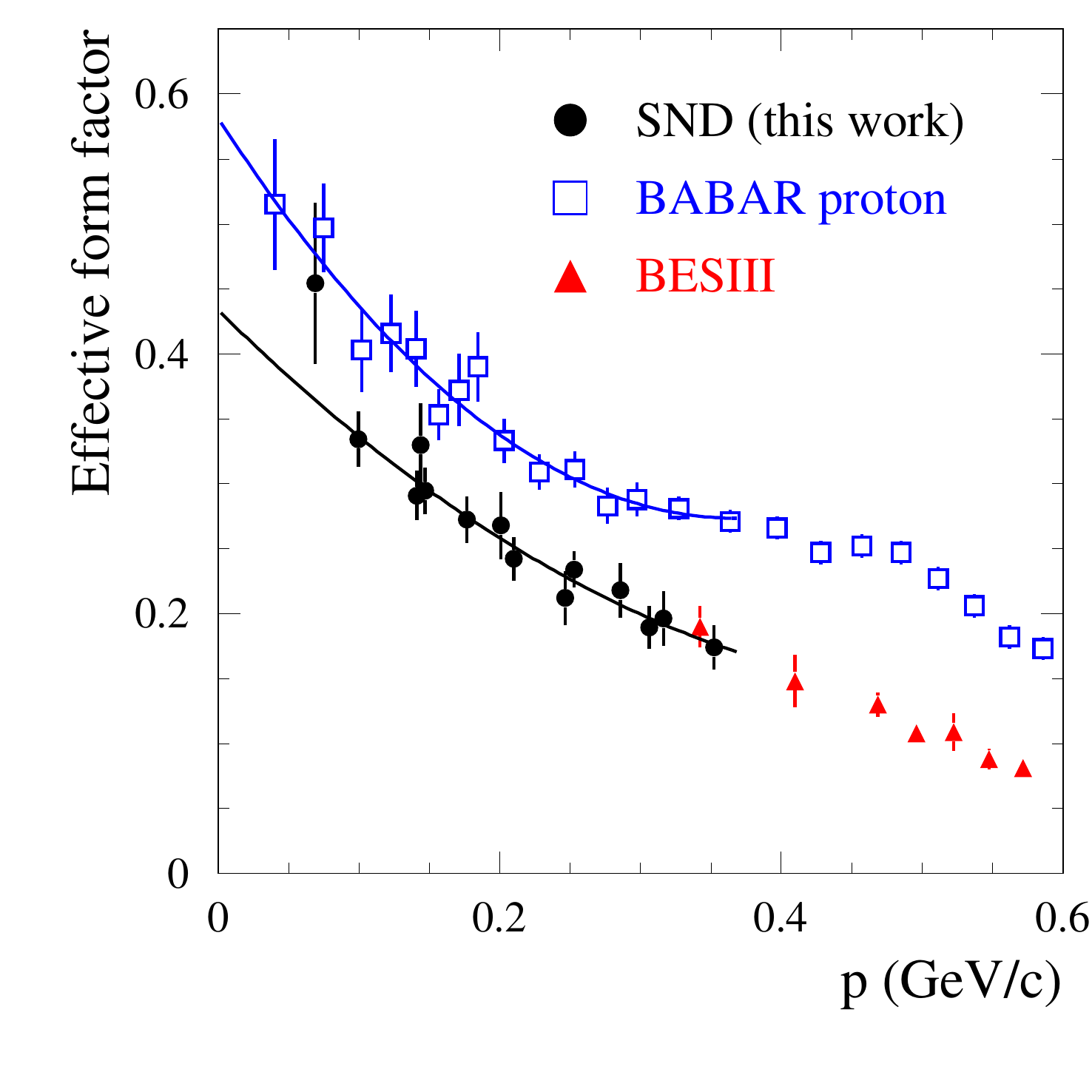}
\caption {  The neutron effective form factor as a function of neutron
momentum obtained in this work compared with the BESIII 
measurements~\cite{BES}, and the proton effective form factor
measured by BABAR~\cite{Babar}. The combined statistical and
systematic uncertainties are shown for all three data sets. The curves 
represent results the fit to the neutron and proton form factor data with 
a second order polynomial.
\label{fig:form}}
\end{figure}

The effective neutron form factor calculated from the measured cross 
section using Eq.~(\ref{eqB2}) is listed in Table \ref{tab:crsect}.  
The form factor as a function of the neutron momentum is shown in 
Fig.~\ref{fig:form} together with the BESIII data~\cite{BES} and the proton 
effective form factor measured by the BABAR experiment~\cite{Babar}.
The curve in Fig.~\ref{fig:form} approximating the SND neutron
form factor is the result of the fit with Model I described above. The
second curve is the result of the fit to the proton form-factor data
with a second-order polynomial. It is seen that Model I can be
successfully applied both for neutron and proton data at momentum region
below 0.35 GeV. In this region the ratio of the proton and neutron
form factors varies from 1.3 to 1.5.

\begin{figure}
\centering
\includegraphics[width=0.47\textwidth]{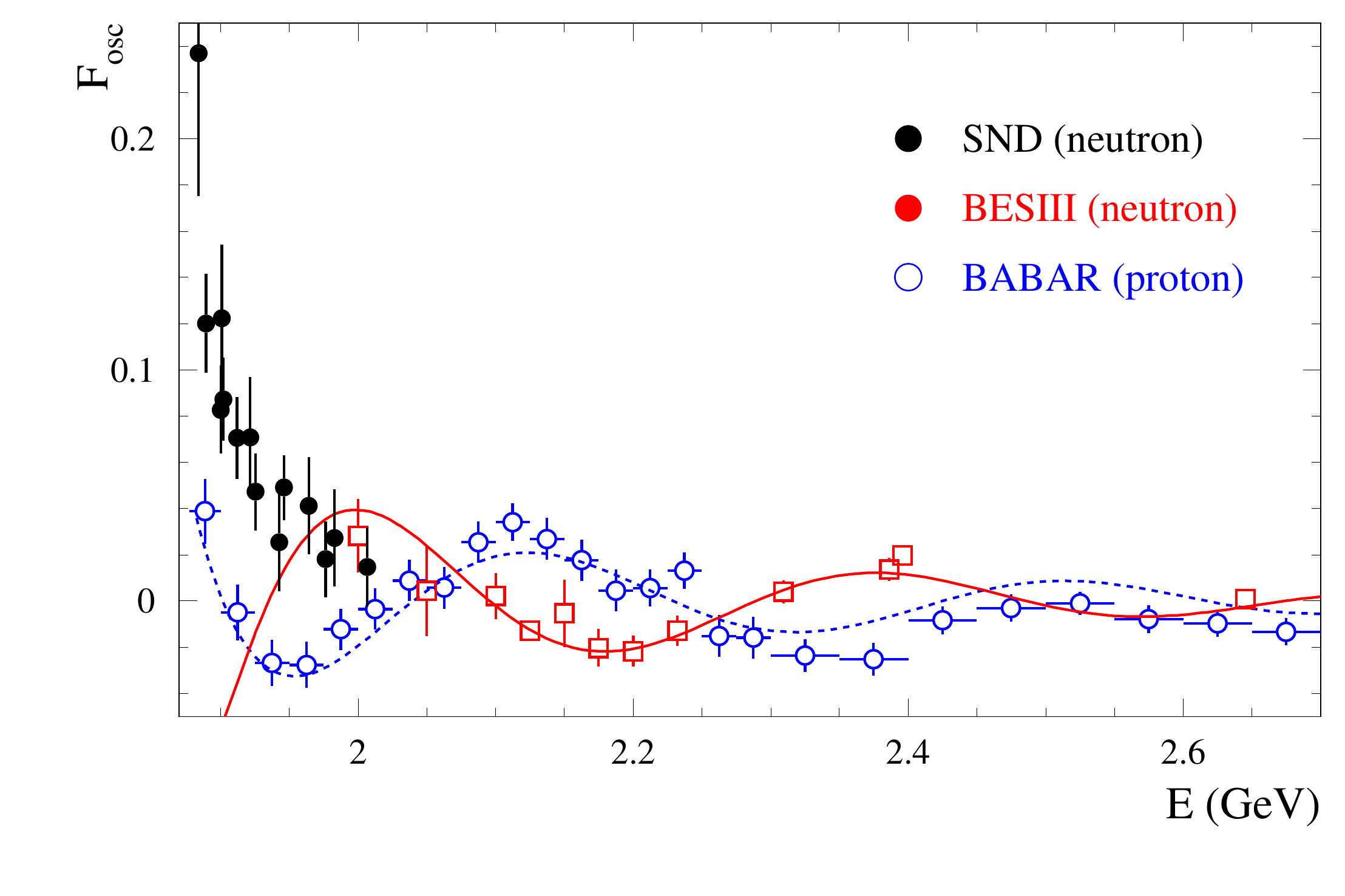}
\caption {The deviation of the proton and neutron effective form factor
data from the dipole formula [see Eq.~(\ref{osc-2})]. The curves are
the result of the simultaneous fit to the BABAR proton and BESIII
neutron
data described in the text.
\label{osc}}
\end{figure}
\begin{figure}
\centering
\includegraphics[width=0.47\textwidth]{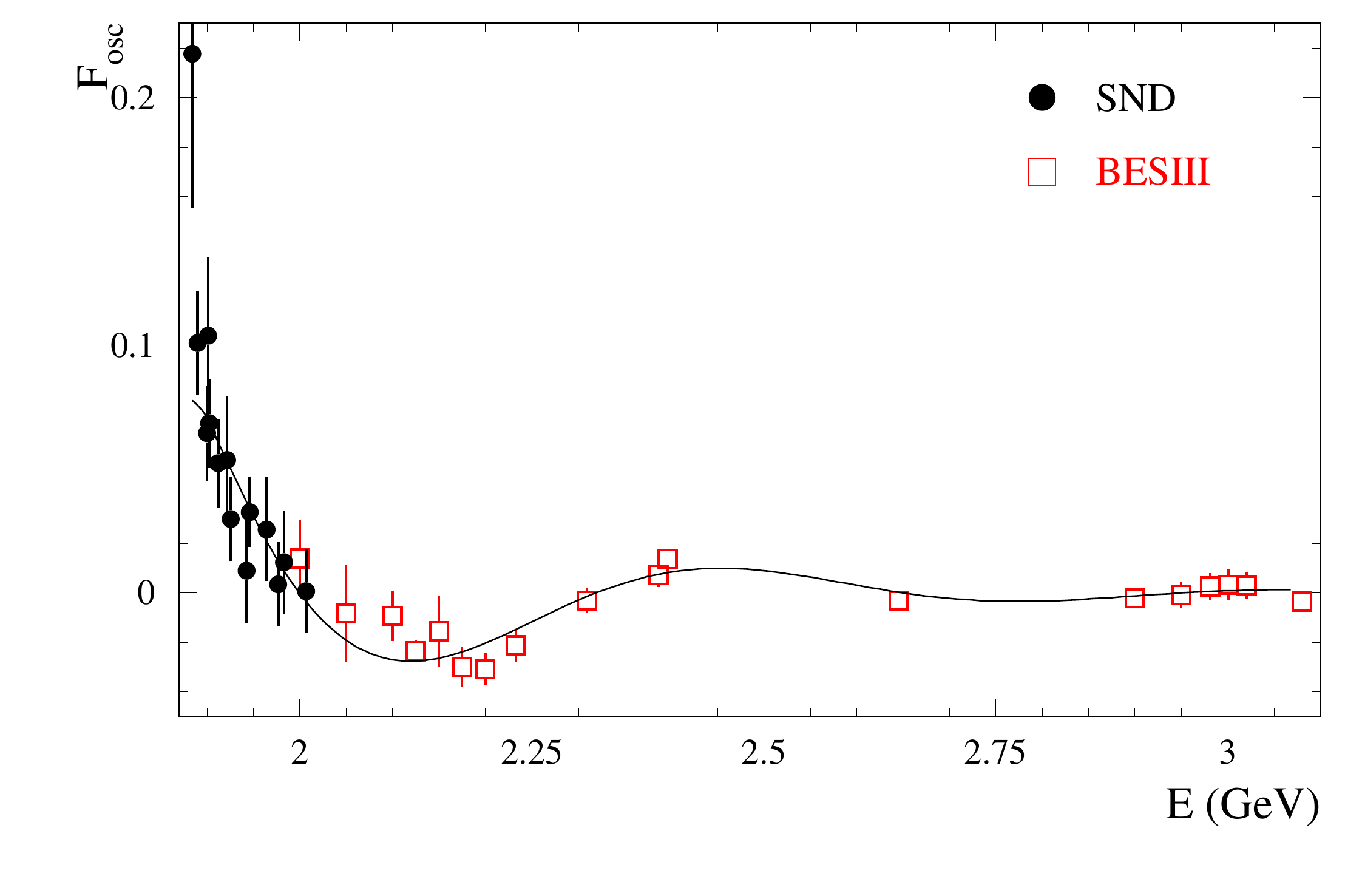}
\caption {The deviation of the neutron effective form factor
data from the dipole formula [see Eq.~(\ref{osc-2})]. The curve is
the result of the fit to the SND and BESIII neutron
data described in the text.
\label{osc1}}
\end{figure}

In Ref.~\cite{osc} a sinusoidal modulation was observed in the proton
effective form factor measured by BABAR~\cite{Babar,Babar1} when
plotting
the data as a function of the proton momentum in the antiproton rest
frame.
These oscillations are seen in Fig.~\ref{osc}, where the difference
between the
BABAR form factor data and a function smooth on the GeV/$c$ scale are
shown.
The latter function~\cite{osc} is obtained using a 2-parameter fit to
the
data in the energy range from the threshold up to 6 GeV.
The same analysis was performed by the BES collaboration for their
neutron
form factor data. The form factor energy dependence was described as
follows:
\begin{eqnarray}
F(s)&=&F_0(s)+F_{\rm osc}(s),\label{osc-1}\\
F_0(s)&=&\frac{{\cal A}_n}{\left[1-s/0.71({\rm
GeV}^2)\right]^2},\label{osc-2}\\
F_{\rm osc}(s)&=&A\exp{(-Bp)}\cos{(Cp + D)},\label{osc-3}\\
p&=&\sqrt{(s/2m_n-m_n)^2-m_n^2}.\label{osc-4}
\end{eqnarray}

At first, the form factor data are fitted by Eq.~(\ref{osc-2}). The
difference $F(s)-F_0(s)$ is plotted in Fig.~\ref{osc}. Then BESIII
performs the simultaneous fit to the BABAR proton and BESIII neutron
data with Eq.~(\ref{osc-3}). The fit parameters $A$, $B$, and $D$ are
different for the proton and neutron data sets, while $C$ is common. The
momentum $p$ for protons is calculated with the substitution $m_n\to m_p$.
The result of this fit is shown in Fig.~\ref{osc}. It is seen that the
model
with a common proton/neutron oscillation frequency $C$ predicts a
specific
energy dependence of the neutron form factor in the energy region
below 2 GeV.
The SND results also plotted in Fig.~\ref{osc} strongly contradict this
prediction. The simultaneous fit
to all three data sets cannot be performed with acceptable quality.
We fit the SND and BES data to Eqs.~(\ref{osc-1})--(\ref{osc-4}). The
result
is shown in Fig.~\ref{osc1}. We obtain a reasonable fit quality
$\chi^2/\nu=31/28$, where $\nu$ is the number of degrees of freedom.
The fitted frequency $C_n=3.3\pm1.7$ GeV$^{-1}$ is significantly lower
than that obtained from the fit to the proton data $C_p=5.6\pm1.9$
GeV$^{-1}$.

\section{Summary \label{sec:summary} }
The experiment to measure $e^+e^-\to n\bar{n}$ cross section has been
carried out with the SND detector at the VEPP-2000 $e^+e^-$ collider
in the energy region from 1884 to 2007 MeV. The measured 
$e^+e^-\to n\bar{n}$ cross varies slowly with energy and is about 0.4 nb 
below 2 GeV. This value is considerably smaller than
the previous measurements of the FENICE~\cite{FENICE} and
SND~\cite{SNND} Collaborations. Near 2 GeV our results agrees 
with the recent BESIII measurement~\cite{BES}. The new SND measurement
supersedes the result of Ref.~\cite{SNND}.

From the measured cross section the neutron effective timelike form factor 
has been extracted. In the energy region under study the ratio of the
proton and neutron effective form factors varies in the range 1.3--1.5.
Using the measured antineutron $\cos\theta$ distribution
the ratio of the electric and magnetic neutron form factors $|G_E|/|G_M|$ 
has been obtained. The results agree with the assumption that $|G_E/G_M|=1$,
but also do not contradict larger values $|G_E/G_M|\approx 1.4$--1.5
observed in the BABAR~\cite{Babar} and BESIII~\cite{BESpp} experiments for the
ratio of the proton form factors near $E=2$ GeV.

{\bf ACKNOWLEDGMENTS}. This work is supported by the Russian
Foundation for Basic Research, grant 20-02-00347 A.


\begin{thebibliography}{1}

\bibitem{FENICE}
A.~Antonelli {\it et al.} (FENICE Collaboration), 
Nucl.\ Phys.\ B {\bf 517}, 3 (1998).
%http://dx.doi.org/10.1016/S0550-3213(98)00083-2

\bibitem{SNND}
M.~N.~Achasov {\it et al.} (SND Collaboration),
Phys.\ Rev.\ D {\bf 517}, 112007 (2014). 
%http://dx.doi.org/10.1103\-/PhysRevD.90.112007

\bibitem{BES} 
M.~Ablikim {\it et al.} (BESIII Collaboration),
Nat.\ Phys.\ {\bf 17},  1200  (2021).
%https://doi.org/10.1038/s41567-021-01345-6

\bibitem{VEPP2k}
P.~Yu.~Shatunov {\it et al.},
Part.\ Nucl.\ Lett.\ {\bf 13}, 995 (2016). 
%http://dx.doi.org/10.1134/S154747711607044X

\bibitem{SNDet1}
M.~N.~Achasov {\it et al.} (SND Collaboration),
Nucl.\ Instrum.\ Meth.\ A {\bf 449}, 125 (2000).
%http://dx.doi.org/10.1016/S0168-9002(99)01302-9
%[hep-ex/9909015]

\bibitem{SNDet2}
V.~M.~Aulchenko {\it et al.},
Nucl. \ Instrum. \ Meth.\ A {\bf 598}, 102 (2009).
http://dx.doi.org/10.1016/\-j.nima.2008.08.099

\bibitem{SNDet3}
A.~Y.~Barnyakov {\it et al.},
Nucl.\ Instrum.\ Meth.\ A {\bf 598}, 163 (2009). 
%http://dx.doi.org/10.1016/\-j.nima.2008.08.018

\bibitem{SNDet4}
V.~M.~Aulchenko {\it et al.},
Nucl.\ Instrum.\ Meth. A {\bf 598}, 340 (2009).
%http://dx.doi.org/10.1016/\-j.nima.2008.08.127

\bibitem{CBS}
E.~V.~Abakumova {\it et al.},
Nucl.\ Insrum.\ Meth.\ A {\bf 744}, 35 (2014). 
%http://dx.doi.org/10.1134/\-S154747711607044X

\bibitem{Annih}
M.~Astrua {\it et al.},
Nucl.\ Phys.\ A  {\bf 697}, 209 (2002). 
%http://dx.doi.org/10.1016/\-S0375-9474(01)01252-0

\bibitem{xi2gam}
A.~V.~Bozhenok {\it et al.},
Nucl.\ Instr.\ Meth.\ A {\bf 379}, 507 (1996).
%http://dx.doi.org/10.1016/0168-9002(96)00548-7

\bibitem{GEANT4}
J.~Allison {\it et al.} (GEANT Collaboration),
Nucl.\ Instr.\ Meth.\ A {\bf 835}, 186 (2016). 
%https://doi.org/10.1016/j.nima.2016.06.125,
%https://geant4-data.web.cern.ch/\-ReleaseNotes/ReleaseNotes4.10.5.html

\bibitem{Timr}
M.~N.~Achasov {\it et al.},
JINST {\bf 10}, T06002 (2015). 
%http://dx.doi.org/10.1088/1748-0221/10/06/T06002

\bibitem{Radcor}
E.~A.~Kuraev and V.~S.~Fadin, 
Sov.\ J.\ Nucl.\ Phys.\ {\bf 41}, 466 (1985).

\bibitem{cmd}
R.~R.~Akhmetshin {\it et al.} (CMD-3 Collaboration),
Phys.\ Lett.\ B {\bf 794}, 64 (2019). 
%http://dx.doi.org/10.1016/j.physletb.2019.05.032

\bibitem{Babar}
J.~P.~Lees {\it et al.} (BABAR Collaboration),
Phys.\ Rev.\ D {\bf 87}, 092005 (2013). 
%http://dx.doi.org/10.1103/\-PhysRevD.87.092005

\bibitem{BESpp}
M.~Ablikim \textit{et al.} (BESIII Collaboration),
Phys.\ Rev.\ Lett. \textbf{124}, 042001 (2020).
%http://dx.doi.org/10.1103/\-PhysRevLett.124.042001

\bibitem{osc}
A.~Bianconi and E.~Tomasi-Gustafsson,
%Periodic interference structures in the timelike proton form factor,''
Phys. Rev. Lett. \textbf{114}, 232301 (2015).
%http://dx.doi.org/10.1103/PhysRevLett.114.232301
%[arXiv:1503.02140 [nucl-th]].

\bibitem{Babar1}
J.~P.~Lees {\it et al.} (BABAR Collaboration),
%Measurement of the $e^+e^- \to p\bar{p}$ cross section in the energy
% range from 3.0 to 6.5 GeV,''
Phys.\ Rev.\ D {\bf 88}, 072009 (2013).
%http://dx.doi.org/10.1103/PhysRevD.88.072009


\end{thebibliography}
\end{document}